\documentclass[aip,jmp,tightenlines,author-year,groupedaddress,nofootinbib]{revtex4-1}

\usepackage[latin1]{inputenc}
\usepackage{graphicx}
\usepackage{amsmath}
\usepackage{subfigure}
\usepackage{verbatim} 

\newcommand{\Cop}{\mathrm{Cop}}
\newcommand{\cop}{\mathrm{cop}}

\begin{document}

\title{Impact of non-stationarity on estimating and modeling empirical copulas of daily stock returns}

\author{Marcel Wollschl{\"a}ger}
\email{marcel.wollschlaeger@uni-due.de}
\affiliation{Faculty of Physics, University of Duisburg-Essen, Lotharstrasse 1, 47057 Duisburg, Germany}
\author{R.~Sch\"afer}
\email{rudi.schaefer@uni-due.de}
\affiliation{Faculty of Physics, University of Duisburg-Essen, Lotharstrasse 1, 47057 Duisburg, Germany}

\begin{abstract}
All too often measuring statistical dependencies between financial time series is reduced to a linear correlation coefficient. However this may not capture all facets of reality.
We study empirical dependencies of daily stock returns by their pairwise copulas. 
Here we investigate particularly to which extent the non-stationarity of financial time series affects both the estimation and the modeling of empirical copulas. 
We estimate empirical copulas from the non-stationary, original return time series and stationary, locally normalized ones. Thereby we are able to explore the empirical dependence structure on two different scales: a \emph{global} and a \emph{local} one. Additionally the asymmetry of the empirical copulas is emphasized as a fundamental characteristic.
We compare our empirical findings with a single Gaussian copula, with a correlation-weighted average of Gaussian copulas, with the K-copula directly addressing the non-stationarity of dependencies as a model parameter, and with the skewed Student's t-copula. 
The K-copula covers the empirical dependence structure on the local scale most adequately, whereas the skewed Student's t-copula best captures the asymmetry of the empirical copula on the global scale.
\end{abstract}

\maketitle

\section{Introduction}
\label{sec:motivation}

The study of empirical dependencies of financial time series is not only important for risk management and portfolio optimization, but it is a prerequisite for a deeper understanding. 
It is common practice to reduce the question about statistical dependence to the study of the Pearson correlation coefficient; the most notable works in this direction include  \cite{markowitz52}, \cite{martens01}, \cite{pelletier05}, \cite{engle06}.
The correlation coefficient, however, only measures the degree of \emph{linear} dependence between two random variables. Non-linear dependencies or more complicated dependence structures are not captured appropriately. 
Here we choose a copula approach to investigate the full statistical dependencies. In contrast to the multivariate distribution which also contains the marginal distributions, the copula, introduced by \cite{Sklar1973}, transforms these marginal distributions to uniform distributions. This allows for a direct study of statistical dependencies. 
Copulas find application in many fields, for example in civil engineering, see \cite{Kilgore2011}, in medicine, see \cite{Onken2009}, in climate and weather research, see \cite{Schoelzel2008} or in the generation of random vectors, see \cite{Strelen2009}. 
In finance, copulas are primarily used in risk and portfolio management, see eg, \cite{Embrechts2001}, \cite{Embrechts2002}, \cite{McNeil2005}, \cite{Rosenberg2006}, \cite{Brigo2010}, \cite{Meucci2011} and \cite{Low2013}. A compact survey of the vast existing and growing literature of copula models provides \cite{Patton2012a} and \cite{Patton2012b} in more detail. For a more general introduction to copulas, the reader is referred to \cite{Joe1997} and \cite{Nelsen2006}.

In all these contexts, easily tractable analytical copulas are used as a building block for the multivariate distributions. Empirical studies which do not assume an analytical dependence structure \emph{a priori} are few and far between, see eg, \cite{Ning2008} and  \cite{Muennix2012}. 
This is where our contribution fits in. We study empirical dependencies of daily returns of S\&P 500 stocks.
To achieve a good statistical significance for the estimation of a full copula, a very large amount of data is necessary. Therefore, we restrict ourselves to the bivariate case and average over many pairwise copulas. In addition, we consider a rather large time horizon.
The consideration of a large time horizon confronts us with two problems: the non-stationarity of the individual time series \emph{and} their dependencies. The former, ie time-varying trends and volatilities, has to be taken into account when we estimate empirical copulas. 
To this end we apply the local normalization to the empirical return time series, which rids them of non-stationary trends and volatilities. The local normalization was introduced by \cite{Schaefer2010} and yields stationary, standard normal distributed time series while preserving dependencies between different time series. It particularly allows us to study the empirical dependence structure on a \emph{local} scale, whereas the copula of original returns reveals the dependence structure on a \emph{global} scale.

We compare our empirical findings with different analytical copulas, in particular addressing the question how far the assumption of a Gaussian dependence structure may carry.
Since we average over many stock pairs, it would be na\"{\i}ve to assume that the resulting average pairwise copula could be described by a single Gaussian copula where only the average correlation enters. And indeed we find rather poor agreement, especially on the global scale. A correlation-weighted average of Gaussian copulas, which takes into account the different pairwise correlations, provides only a slightly better description.
We have to consider the time-varying dependencies as well. These can be addressed  by an ensemble average over random correlations, see \cite{Schmitt2013}. This ansatz yields good agreement for multivariate return distributions. Here we derive the pairwise K-copula resulting from this random matrix approach and compare it with our empirical findings. Overall, we find rather good agreement, especially on the local scale. However, the empirical copulas show an asymmetry in the tail dependence which our model cannot account for. 
As a model that allows for an asymmetry in the dependence structure the skewed Student's t-distribution, first introduced by \cite{Hansen1994}, and its corresponding copula, introduced by \cite{Demarta2005}, are steadily gaining popularity in particular with regard to credit risk management, see eg, \cite{Hu2006}, \cite{Dokov2008}, and the study of asymmetric dependencies, see eg, \cite{Sun2008}. On the global scale where the asymmetry is more pronounced, the skewed Student's t-copula yields improved agreement with our empirical copulas.

The paper is structured as follows. In section \ref{sec:theory}, we present briefly our data set and the theoretical background for the local normalization, correlations and copulas. Additionally, we derive the K-copula. We present our results in section \ref{sec:results} and conclude in section \ref{sec:conclusion}.

\section{Data set and theoretical background}
\label{sec:theory}

\subsection{Prices and returns}
\label{sec:data}
In our empirical study, we consider the daily closing prices of stocks in the S\&P 500 stock index, which have been adjusted for splits and dividends. Our observation period ranges from 03/01/2006 to 12/31/2012. The data are obtained from \texttt{http://finance.yahoo.com/}. From the price time series we calculate the returns---later referred to as \emph{original} returns---as
\begin{align}
r_k(t) = \frac{S_k(t + \Delta t) - S_k(t)}{S_k(t)} 
\label{eq:returns}
\end{align}
where $S_k(t)$ is the price at time $t$ of a stock $k$ and $\Delta t=1$ trading day. We finally arrive at $T=1760$ daily returns for $K=460$ stocks which were continuously traded and listed in the S\&P 500 during the entire observation period. It is well known that original returns show strongly non-stationary behavior.

\subsection{Local normalization}
\label{sec:locnorm}
To correct for non-stationary trends and volatilities in the original returns, we employ a method called local normalization, introduced by \cite{Schaefer2010}. 
Here the locally normalized return is defined as 
\begin{align}
\rho_k(t) = \frac{r_k(t)-\mu_k(t)}{\sigma_k(t)}
\end{align}
with the local average $\mu_k(t)$ and the local volatility $\sigma_k(t)$ of a stock $k$ which are both estimated over a time window of the preceding 13 trading days. As has been shown in \cite{Schaefer2010}, the interval of 13 trading days for the local average and volatility yields the best approximation of stationary, standard normal distributed time series.

\subsection{Correlations}
\label{sec:correlations}
To measure the statistical dependence of two random variables $X, Y$, one often considers the Pearson correlation coefficient, \begin{align}
C_{X,Y} = \frac{\mathrm{Cov}(X,Y)}{\sigma_X\,\sigma_Y}
\label{eq:corr_def}
\end{align}
where $\mathrm{Cov}(X,Y)$ is the covariance of both random variables and $\sigma_X$, $\sigma_Y$ are their respective standard deviations. In our case the random variables $X, Y$ are returns $r_k, r_l$. 
Non-stationarity of the individual time series is a fundamental problem for the estimation of  correlation coefficients. For return time series, the time-varying volatilities lead to estimation errors in equation \eqref{eq:corr_def}, see eg, \cite{Schaefer2010}, \cite{Muennix2012}. This can be taken into account by the local normalization introduced above. 
As already mentioned in the introduction, the correlation coefficient only measures the linear dependence between two random variables. 
Therefore, it is not well-suited to measure arbitrary dependencies. For this purpose we consider copulas in the following section.

\subsection{Definition of copulas}
\label{sec:copula}
In our short introduction to copulas we shall confine ourselves to the bivariate case, since we only study empirical \emph{pairwise} copulas. 
The full information about the statistical dependence of two random variables is certainly contained in their bivariate distribution. However, it is difficult to compare bivariate distributions, if the marginal distributions are not the same for all random variables. To achieve comparability we can transform each marginal distribution to a uniform distribution. The bivariate distribution of the thusly transformed variables is then called a copula.

We consider two continuous random variables $X,Y$ with a joint probability density function $f_{X,Y}(x,y)$. Then the joint cumulative distribution function is given by
\begin{align}
F_{X,Y}(x,y) = \int\limits_{-\infty}^x\mathrm dx' \int\limits_{-\infty}^y\mathrm dy'\,{f_{X,Y}(x',y')}
\end{align}

We denote the marginal distribution functions by $F_X(x)$, $F_Y(y)$ and the corresponding probability density functions by $f_X(x) = \frac{\mathrm d}{\mathrm dx} F_X(x)$, $f_Y(y) = \frac{\mathrm d}{\mathrm dy} F_Y(y)$. 
According to \cite{Sklar1973}, we define the copula by
\begin{align}
F_{X,Y}(x,y) &= \Cop_{X,Y}(F_X(x),F_Y(y)) \nonumber\\
\Leftrightarrow \quad \Cop_{X,Y}(u,v) &= F_{X,Y}(F_X^{-1}(u),F_Y^{-1}(v)) 
\end{align}
where we used the transformation $u=F_X(x)$, $v=F_Y(y)$, and $F_X^{-1},F_Y^{-1}$ indicate the inverse marginal distribution functions, also called quantile functions. According to the usual definition of a probability density function, the copula density is defined by the partial derivatives with respect to $u$ and $v$:
\begin{align} \label{eq:RMTCop}
\cop_{X,Y}(u,v) &= \frac{\partial^2 \Cop_{X,Y}(u,v) }{\partial u \partial v}
= \left.\frac{\partial^2 F_{X,Y}(x,y)}{\partial x\partial y}\right|_{\substack{x=F_X^{-1}(u)\\y=F_Y^{-1}(v)}}\frac{\partial F_X^{-1}(u)}{\partial u}\frac{\partial F_Y^{-1}(v)}{\partial v} \\
&= \left.\frac{f_{X,Y}(x,y)}{f_X(x) f_Y(y)}\right|_{\substack{x=F_X^{-1}(u)\\y=F_Y^{-1}(v)}} \nonumber
\end{align}
where we used the inverse function rule in the last step.

\subsection{Empirical pairwise copula}
\label{sec:empcoptheory}
In order to calculate the empirical pairwise copula of two return time series $r_k,r_l$,  
we have to transform their marginal distributions to uniformity. We then refer to the transformed time series as $u=u_k$, $v=u_l$. To achieve uniform distributed time series, we utilize the empirical distribution function \begin{align}
u_k(t) = F_k(r_k(t)) 
= \frac{1}{T}\sum_{\tau=1}^T{\mathbf{1}\{r_k(\tau) \leq r_k(t)\}} - \frac{1}{2T} 
\label{eq:u}
\end{align}
where $\mathbf{1}$ is the indicator function. 
The term $-\frac12$ in definition \eqref{eq:u} simply ensures that all values $u_k(t)$ lie strictly in the interval (0,1). Finally, we arrive at the empirical pairwise copula density as a two-dimensional histogram of data pairs $(u_k(t),u_l(t))$.

\subsection{Bivariate Gaussian copula}
\label{sec:gausscopula}
\begin{figure}[htbp]
	\centering
		\includegraphics[width=0.49\textwidth]{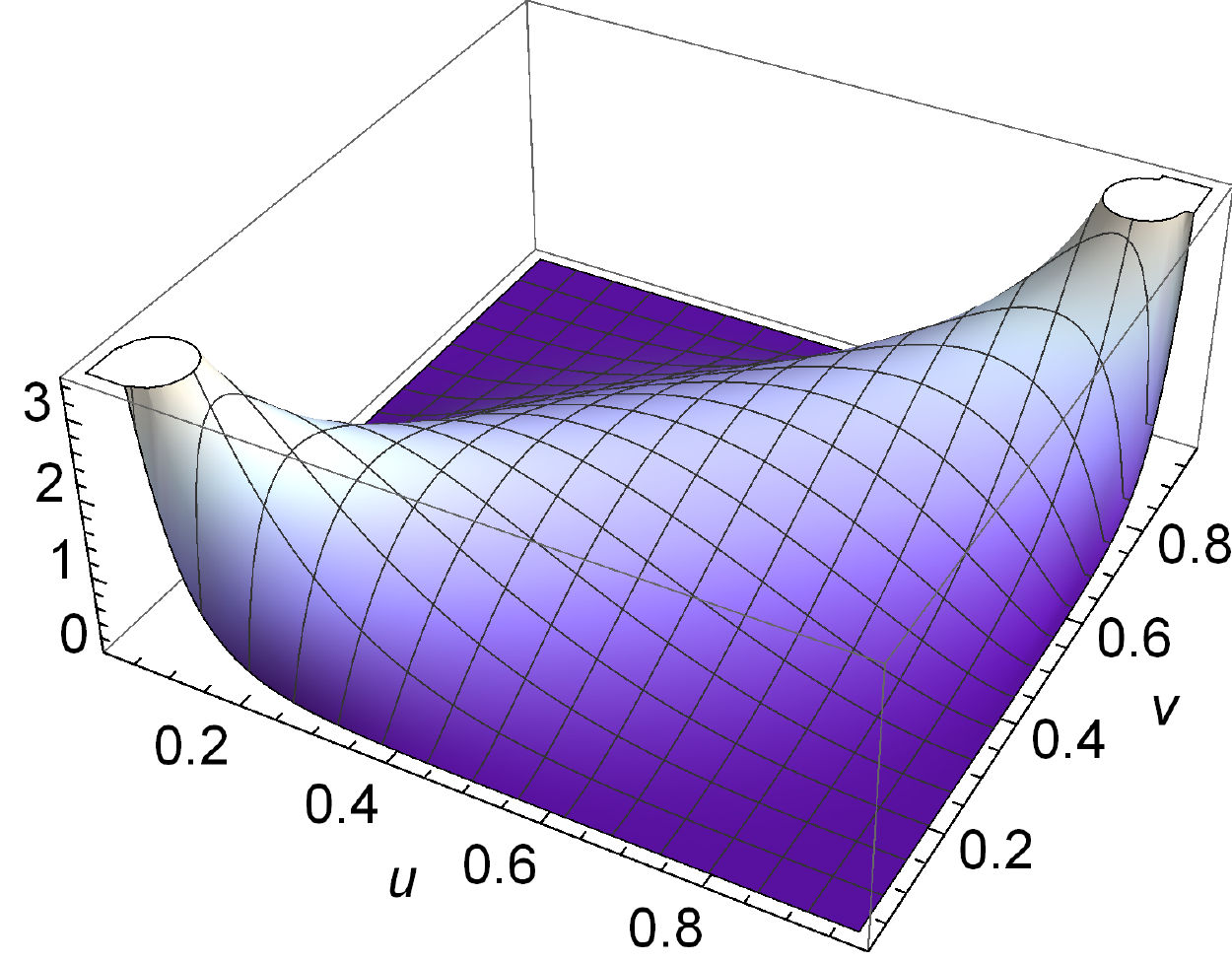}
		\includegraphics[width=0.49\textwidth]{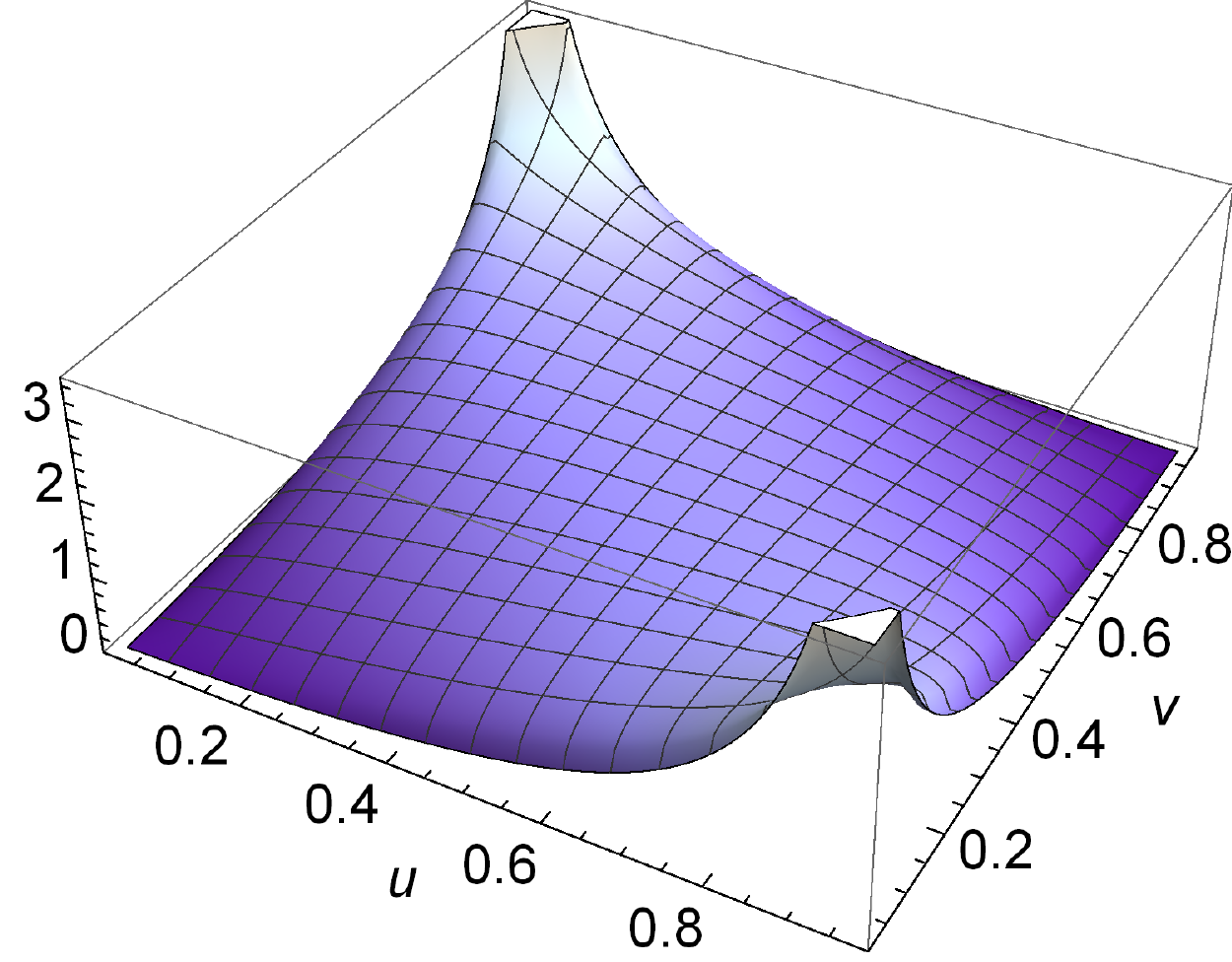}
	\caption{Gaussian copula density $\cop_c^\text G(u,v)$ for different correlations $c$, left: $c=0.8$, right: $c=-0.5$.}
	\label{fig:gausscoptheo}
\end{figure}

The Gaussian copula is the dependence structure that arises from a normal distribution. We consider two random variables $X,Y$ which follow a joint normal distribution with correlation coefficient $c$. The bivariate Gaussian copula is then given by
\begin{align}
\Cop_c^\text G(u,v) = \Phi_c(\Phi^{-1}(u),\Phi^{-1}(v)) 
\end{align}
where $\Phi_c$ describes the cumulative distribution function of a bivariate standard normal distribution and $\Phi^{-1}$ the inverse cumulative distribution function of a univariate standard normal distribution:
\begin{align}
\Phi_c(x,y) &= \int\limits_{-\infty}^x\mathrm dx' \int\limits_{-\infty}^y\mathrm dy'\,{\frac{1}{2\pi\sqrt{1-c^2}}\exp\left(-\frac{(x')^2-2cx'y'+(y')^2}{2(1-c^2)}\right)} \\
\Phi(x) &= \int\limits^x_{-\infty} \frac{1}{\sqrt{2\pi}}\exp\left(-\frac{(x')^2}{2}\right)\mathrm dx' 
\end{align}
Partial derivation yields the bivariate Gaussian copula density
\begin{align}
\cop_c^\text G(u,v) &= \frac{1}{\sqrt{1-c^2}} \\
& \quad \times \exp\left(-\frac{c^2\Phi^{-1}(u)^2-2c\,\Phi^{-1}(u)\Phi^{-1}(v)+c^2\Phi^{-1}(v)^2}{2(1-c^2)}\right)  \nonumber
\end{align}
In figure \ref{fig:gausscoptheo}, we exemplarily show two Gaussian copula densities $\cop_c^\text G(u,v)$ for a positive and a negative correlation $c$, respectively. Here we can distinguish the inherently symmetric character as a fundamental property. Gaussian copula densities with positive correlations describe strong dependencies between events in the same tail of each marginal distribution, while negative correlations describe strong dependencies between events in the opposite tails.

\subsection{Bivariate K-copula}
\label{sec:RMT}
Correlations between financial time series are time-dependent, see eg, \cite{Engle2002}, \cite{Tse2002}, \cite{Cappiello2006}, \cite{Schaefer2010} and \cite{Muennix2012}. To take this empirical fact into account, we consider a random matrix model, introduced by \cite{Schmitt2013}, which assumes for each return vector $\mathbf r(t)=(r_1(t),\dots,r_K(t))$ at each time $t=1,\dots,T$ a multivariate normal distribution
\begin{align}
g(\mathbf r(t),\Sigma_t) = \frac{1}{\sqrt{\det(2\pi \Sigma_t)}}\exp\left(-\frac12 \mathbf r^\dagger(t) \Sigma^{-1}_t \mathbf r(t)\right) 
\end{align}
with a \emph{time-dependent} covariance matrix $\Sigma_t$. Each time-dependent covariance matrix $\Sigma_t$ is then modeled by a random matrix $\Sigma_t\rightarrow AA^\dagger/N$ drawn from a Wishart distribution, see \cite{Wishart1928}. To this end we choose a multivariate normal distribution for the matrix elements of $A$,
\begin{align}
w(A,\Sigma) = \frac{1}{\det^{N/2}(2\pi\Sigma)}\exp\left(-\frac12 \mathrm{tr}A^\dagger\Sigma^{-1}A\right) 
\end{align}
which fluctuate around the empirical average covariance matrix $\Sigma=\langle\Sigma_t\rangle_{t=1,\dots,T}$. The distribution of a sample of return vectors $\mathbf r$ is then given by a multivariate normal-Wishart mixture: We are averaging a multivariate normal distribution over an ensemble of Wishart-distributed covariances which finally results in a K-distribution,
\begin{align} \label{eq:Bessel}
\langle g \rangle(\mathbf r,\Sigma,N) &= \int{\mathrm d[A]\,g\left(\mathbf r,\frac1N AA^\dagger\right)w(A,\Sigma)} \\
&= \frac{1}{2^{N/2+1}\Gamma(N/2)\sqrt{\det(2\pi\Sigma/N)}} \nonumber \\
& \qquad \frac{\mathcal{K}_{(K-N)/2}\left(\sqrt{N \mathbf r^\dagger \Sigma^{-1} \mathbf r}\right)}{\sqrt{N \mathbf r^\dagger \Sigma^{-1} \mathbf r}^{(K-N)/2}} \nonumber \\
&= \frac{1}{(2\pi)^K\Gamma(N/2)\sqrt{\det\Sigma}} \nonumber \\
& \qquad \int\limits_0^\infty{\mathrm dz\,z^{\frac{N}{2}-1}\mathrm e^{-z}\sqrt{\frac{\pi N}{z}}^K\exp\left(-\frac{N}{4z} \mathbf r^\dagger \Sigma^{-1} \mathbf r\right)}  \nonumber
\end{align}
where $\int\mathrm d[A]$ denotes the integral over all matrix elements of $A$, and $\mathcal{K}_\lambda$ is the modified Bessel function of the second kind of order $\lambda$. The K-distribution \eqref{eq:Bessel} contains only two parameters: the empirical average covariance matrix $\Sigma$ and a free parameter $N$ which characterizes the fluctuations of  covariances around the empirical average $\Sigma$. In this manner, the empirically observed non-stationarity of covariances enters directly into the random matrix model. Note that the K-distribution \eqref{eq:Bessel} is a special case of the multivariate generalized hyperbolic distribution, see eg, \cite{McNeil2005}, \cite{Schmidt2006}, \cite{Aas2006}, \cite{Necula2009}, \cite{Socgnia2014}, \cite{Vilca2014} and \cite{Browne2014}.

\begin{figure}[htbp]
	\centering
		\includegraphics[width=0.49\textwidth]{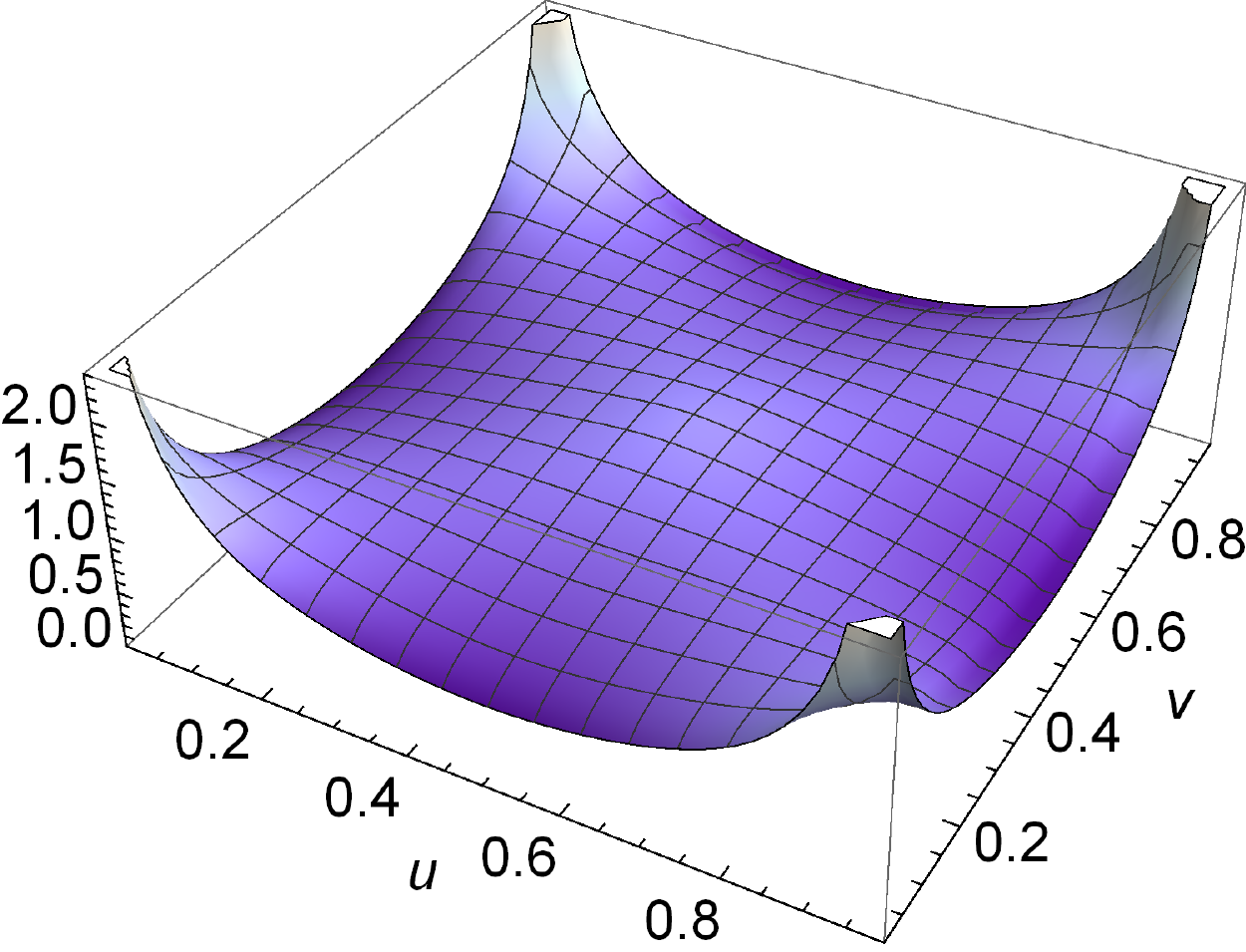}
		\includegraphics[width=0.49\textwidth]{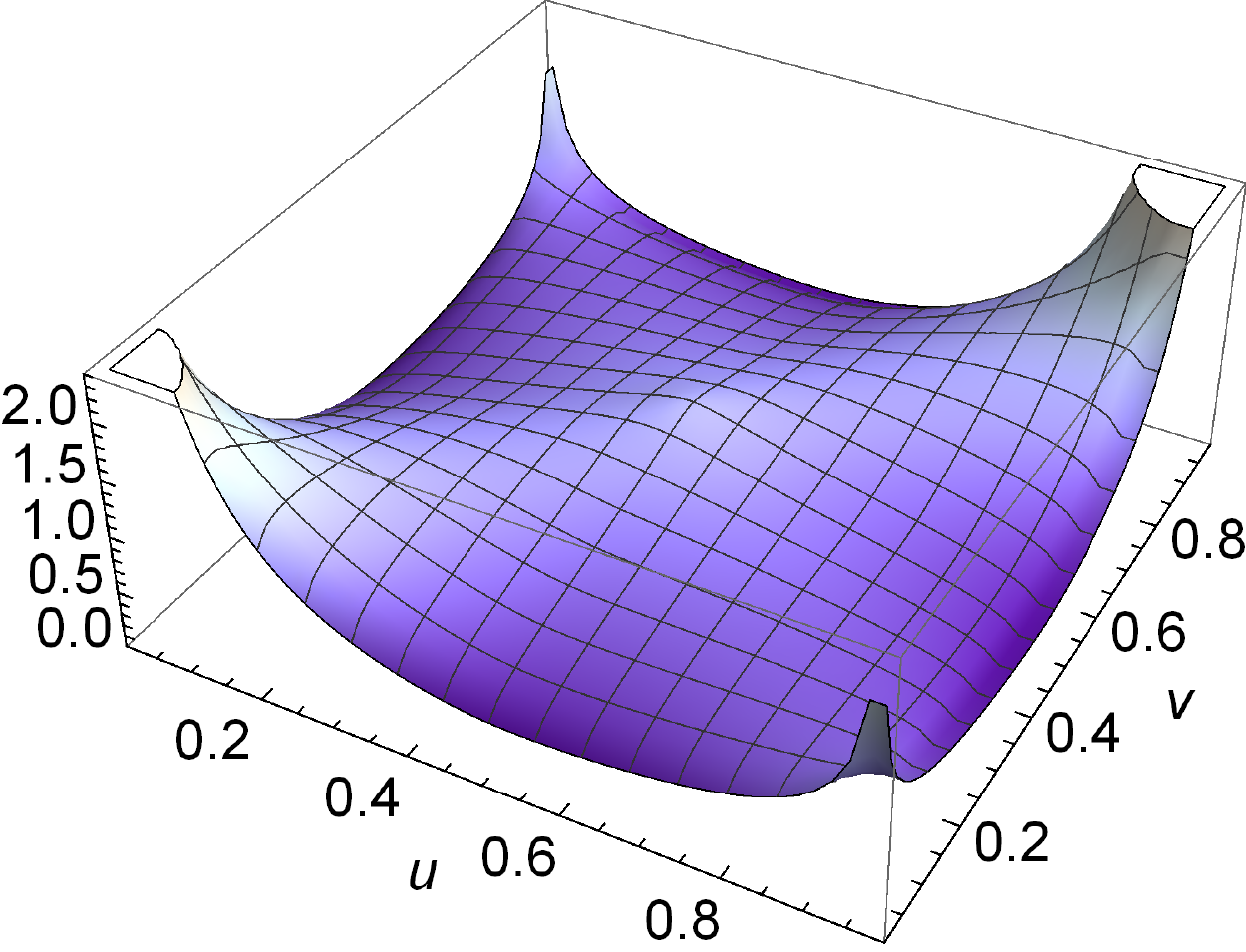}
	\caption{K-copula densities $\cop_{c,N}^\text K(u,v)$ for average correlation $c$ and parameter $N$, left: $c=0$, $N=5$, right: $c=0.2$, $N=4$.}
	\label{fig:copuncorr}
\end{figure}

By considering the K-distribution \eqref{eq:Bessel} for the bivariate case, we are now deriving the bivariate K-copula density via equation \eqref{eq:RMTCop}. In our case, the probability density function is the bivariate K-distribution, $f_{X,Y}(x,y)=\langle g \rangle(\mathbf r,\Sigma,N)$ with $K=2$. Since the copula density is independent of the marginal distributions, we can choose the standard deviations $\sigma_X = \sigma_Y = 1$, leading to the covariance matrix 
\begin{align}
\Sigma =
\begin{pmatrix}
\sigma_X^2 & \sigma_X \sigma_Y c\\
\sigma_X \sigma_Y c & \sigma_Y^2
\end{pmatrix}
=
\begin{pmatrix}
1 & c\\
c & 1
\end{pmatrix} 
\label{eq:sigma}
\end{align}
which is now nothing but a correlation matrix with the empirical average correlation coefficient $c$. Thus, the parameter $N$ simply characterizes the fluctuation strength of the correlations around their mean value. The smaller $N$, the larger the fluctuations. In the limit $N \to \infty$ the fluctuations vanish and we arrive at a normal distribution and a Gaussian copula, respectively.
The marginal probability density functions of the K-distribution \eqref{eq:Bessel} are identical, $f_X(\cdot)=f_Y(\cdot)$, with
\begin{align}
f_X(x) = \int\limits_{-\infty}^\infty{\mathrm dy\,f_{X,Y}(x,y)} = \frac{1}{\Gamma(N/2)}\int\limits_0^\infty{\mathrm dz\,z^{\frac{N}{2}-1}\mathrm e^{-z}\sqrt{\frac{N}{4\pi z}}\exp\left(-\frac{N}{4z}x^2\right)} 
\end{align}
and the marginal cumulative distribution functions, $F_X(\cdot)=F_Y(\cdot)$, with
\begin{align}
F_X(x) &= \int\limits_{-\infty}^x\mathrm d\xi\,f_X(\xi) \\
&= \frac{1}{\Gamma(N/2)}\int\limits_0^\infty{\mathrm dz\,z^{\frac{N}{2}-1}\mathrm e^{-z}\int\limits_{-\infty}^x\mathrm d\xi\,\sqrt{\frac{N}{4\pi z}}\exp\left(-\frac{N}{4z}\xi^2\right)}  \nonumber
\end{align}
have to be calculated and inverted numerically. Insertion into equation \eqref{eq:RMTCop} then yields the K-copula density $\cop_{c,N}^\text K(u,v)$. For illustration, figure \ref{fig:copuncorr} shows the K-copula density for an average correlation $c=0$ and $N=5$ on the one hand, and for $c=0.2$ and $N=4$ on the other hand. We observe that the K-copula density contains both positive and negative correlations in contrast to the Gaussian copula density. In particular negative correlations are covered as well if the average correlation $c$ is positive but the fluctuations around it are large enough. However, the K-copula density is still symmetric with respect to the facing corners.

\subsection{Bivariate skewed Student's t-copula}
\label{sec:skewedtcop}
We consider a bivariate skewed Student's t-distributed random variable $\mathbf Z=(X,Y)$ represented by the normal mean-variance mixture
\begin{align}
\mathbf Z = \boldsymbol\mu + \boldsymbol\gamma W + \mathbf Q \sqrt W
\label{eq:skewedt}
\end{align}
where $\boldsymbol\mu=(\mu_1, \mu_2)$ is a location parameter vector, $W \sim IG(\nu/2,\nu/2)$ follows an inverse gamma distribution with $\nu$ degrees of freedom, $\mathbf Q \sim N(0,\Sigma)$ is independent of $W$ and drawn from a bivariate normal distribution with zero-mean and covariance matrix $\Sigma$ and $\boldsymbol\gamma = (\gamma_1, \gamma_2)$ is the skewness parameter vector. Since we are aiming at deriving a skewed Student's t-copula density from its corresponding distribution following equation \eqref{eq:RMTCop}, we are able to drop the location parameter vector $\boldsymbol\mu$ in the stochastic representation \eqref{eq:skewedt}, ie $\boldsymbol\mu=\mathbf0$, and set the standard deviations to one such that $\Sigma$  conforms to the correlation matrix \eqref{eq:sigma} again. In this case, the bivariate skewed Student's t probability density function reads
\begin{align}
f_\mathbf Z(\mathbf z) = \frac{1}{2^{\nu/2} \Gamma(\nu/2) \pi\nu \sqrt{\det \Sigma}}
\frac{\exp(\mathbf z^\dagger \Sigma^{-1} \boldsymbol\gamma)}{(1+\frac{\mathbf z^\dagger \Sigma^{-1} \mathbf z}{\nu})^{\nu/2+1}}
\frac{\mathcal K_{\nu/2+1}(\sqrt{(\nu + \mathbf z^\dagger \Sigma^{-1} \mathbf z) \boldsymbol\gamma^\dagger \Sigma^{-1} \boldsymbol\gamma})}{(\sqrt{(\nu + \mathbf z^\dagger \Sigma^{-1} \mathbf z) \boldsymbol\gamma^\dagger \Sigma^{-1} \boldsymbol\gamma})^{-(\nu/2+1)}}
\end{align}
where $\mathcal{K}_\lambda$ is the modified Bessel function of the second kind of order $\lambda$. The univariate marginal probability density functions are identical, $f_X(\cdot) = f_Y(\cdot)$, with
\begin{align}
f_X(x) = \frac{1}{2^{(1-\nu)/2} \Gamma(\nu/2) \sqrt{\pi\nu} \sqrt{\det \Sigma}}
\frac{\exp(x \gamma_1)}{(1+\frac{x^2}{\nu})^{(\nu+1)/2}}
\frac{\mathcal K_{(\nu+1)/2}(\sqrt{(\nu + x^2) \gamma_1^2})}{(\sqrt{(\nu + x^2) \gamma_1^2})^{-(\nu+1)/2}}
\end{align}
and the marginal cumulative distribution functions, $F_X(\cdot)=F_Y(\cdot)$, with
\begin{align}
F_X(x) &= \int\limits_{-\infty}^x\mathrm d\xi\,f_X(\xi)
\end{align}
have to be calculated and inverted numerically. Bringing it all together corresponding to equation \eqref{eq:RMTCop} with $f_{X,Y}(x,y) = f_\mathbf Z(\mathbf z)$ then yields the bivariate skewed Student's t-copula density $\cop_{c,\nu,\boldsymbol\gamma}^t(u,v)$. In the limit $\nu\to\infty$ we receive the Gaussian copula density.

\section{Empirical results and model comparison}
\label{sec:results}

In section \ref{sec:empcop} we present the empirical copula densities for original returns and locally normalized returns, respectively. In section \ref{sec:asymmetry} we discuss the asymmetry in the tails of the empirical copula densities in more detail. We then compare the empirical copulas to the Gaussian copula with mean correlation in section \ref{sec:avggausscop}, to a correlation-weighted Gaussian copula in section \ref{sec:corrweigausscop}, to the K-copula in section \ref{sec:kcop} and to the skewed Student's t-copula in section \ref{sec:studcop}.

\subsection{Empirical copula densities}
\label{sec:empcop}
\begin{figure}[htbp]
	\centering
		\includegraphics[width=0.49\textwidth]{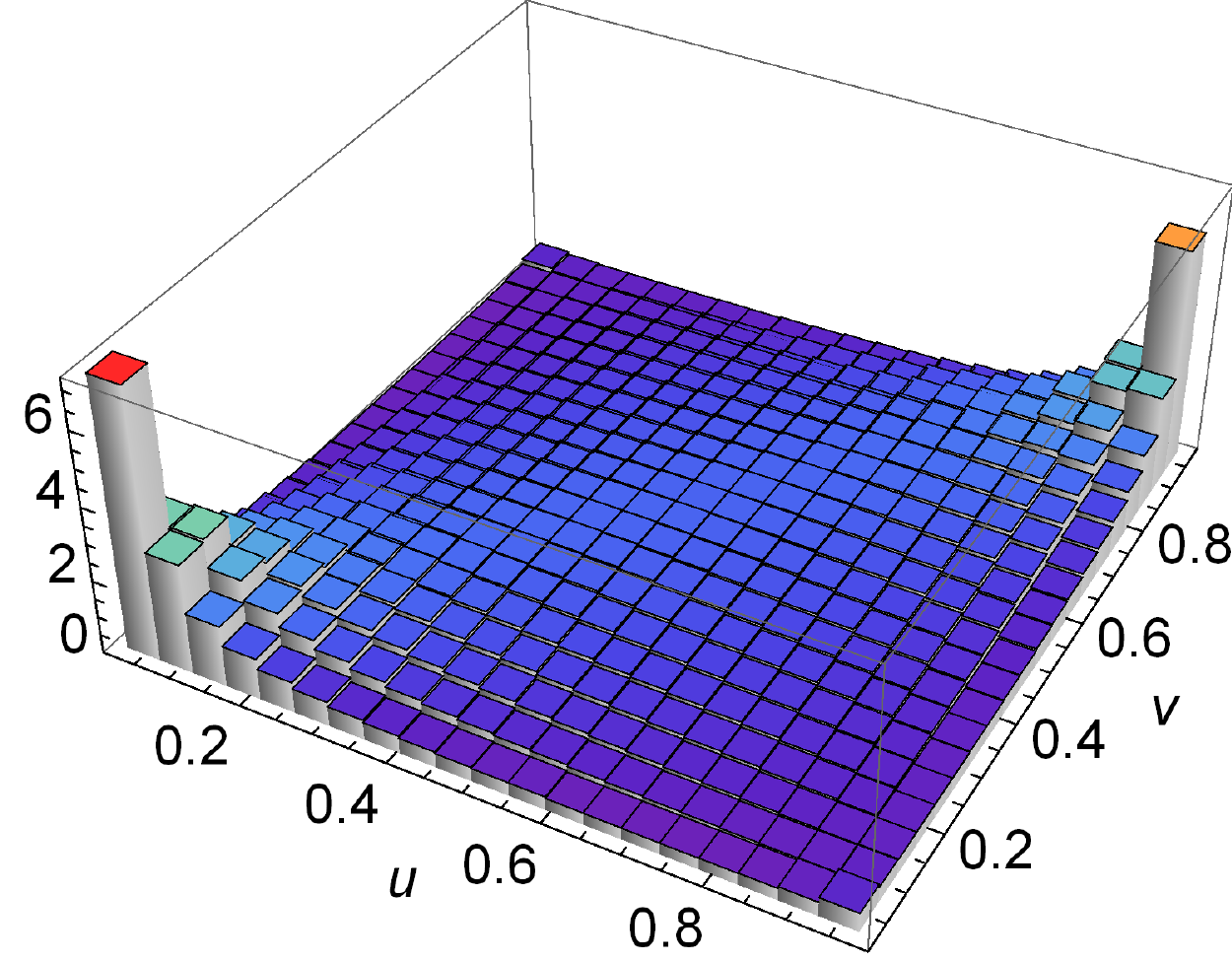}\\
		\includegraphics[width=0.49\textwidth]{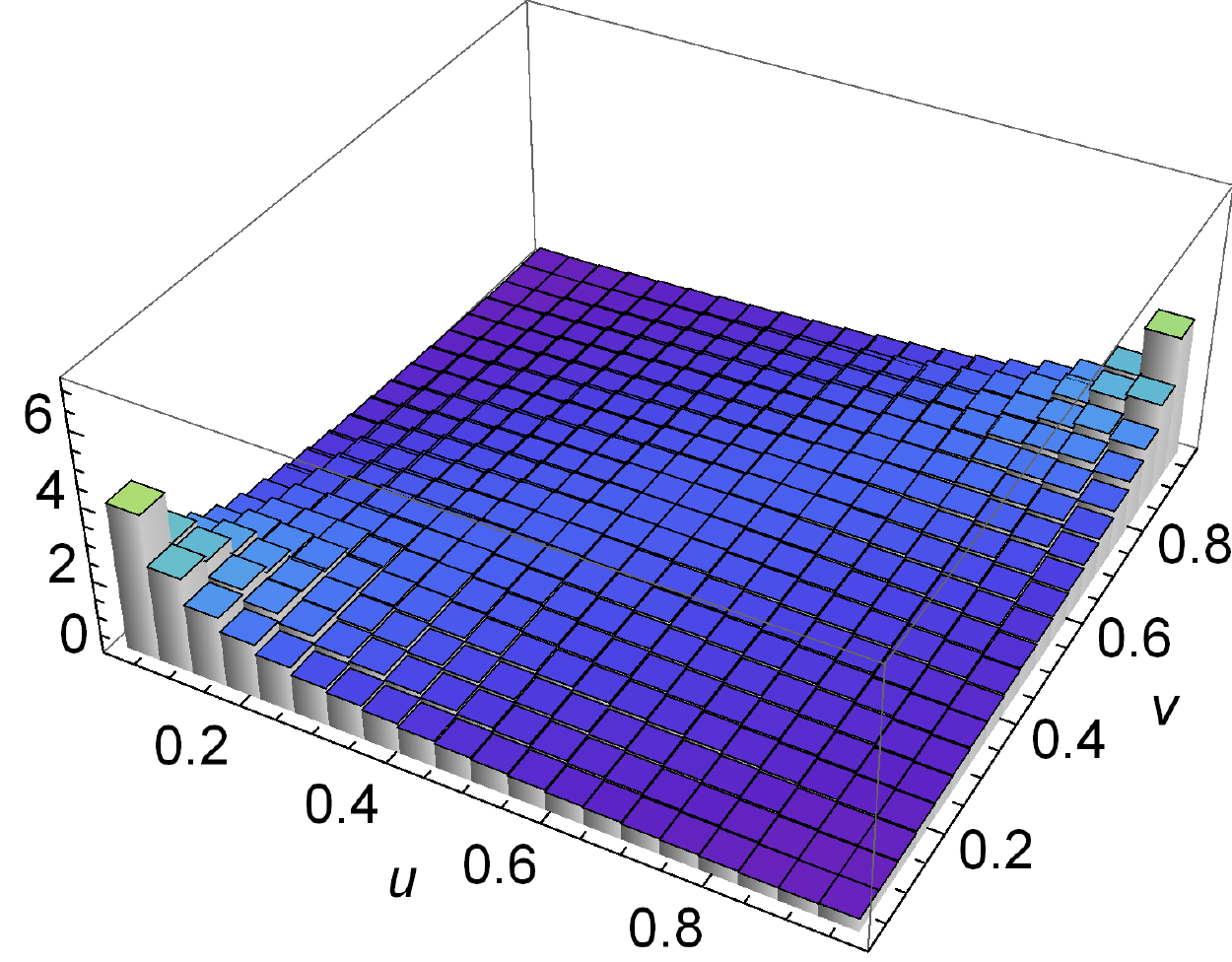}
	\caption{
Empirical pairwise copula densities, top: for original returns, $\cop^\text{(glob)}(u,v)$, bottom: for locally normalized returns, $\cop^\text{(loc)}(u,v)$.
}
	\label{fig:empcop}
\end{figure}

We consider empirical pairwise copula densities averaged over all $K(K-1)/2$ stock pairs
\begin{align}
\cop(u,v) = \frac{2}{K(K-1)} \sum_{k,l=1, \ l>k}^{K} \cop_{k,l}(u,v) \label{equ:avgempcop}
\end{align}
where $\cop_{k,l}(u,v)$ is calculated as a two-dimensional histogram of the data pairs $(u_k(t),u_l(t))$, according to equation \eqref{eq:u}. For the bin size of these histograms we choose $\Delta u=\Delta v=1/20$. In the following, we denote the empirical copula density: $\cop^\text{(glob)}(u,v)$ for original returns, and $\cop^\text{(loc)}(u,v)$ for locally normalized returns. 
As mentioned above, the main difference between these two cases is the consideration of non-stationarity: the original returns exhibit time-varying trends and volatilities; thus, one might argue, the return time series are non-stationary.
The locally normalized returns, on the other hand, show stationary behavior.
We compare the results for both cases because each has its merits. The copula for the original returns provides the statistical dependence over the full time horizon, ie on a \emph{global} scale, while the copula for the locally normalized returns reveals the statistical dependence on a \emph{local} scale. 
In figure \ref{fig:empcop}, we compare the empirical copula densities. On first sight, the dependence structure seems very similar for both cases. Deviations exist mainly in the corners. 
Overall, the statistical dependence is preserved rather well when we strip the time series from time-varying trends and volatilities. 
For the local normalization, \cite{Schaefer2010} showed that correlations are preserved under this procedure. Apparently, this also holds for the copula to some degree.
For the original returns, the two peaks in the corners are higher. This can be explained as follows: 
The returns of periods with high volatility are more likely to end up in the lowest or highest quantile, ie $u_k(t)$ and $u_l(t)$ are close to 0 or 1. Therefore, periods with high volatility contribute strongly to the corners of this copula. Since high volatility typically coincides with strong correlations in the market, the corners exhibit a stronger dependence.

Qualitatively, the empirical results resemble Gaussian copula densities except for the corners, where an asymmetry is observed. This empirical asymmetry implies that large negative returns of two stocks show stronger dependence than large positive ones. Although the asymmetry can be captured neither by the Gaussian copula or a Gaussian mixture nor by the K-copula, we are going to investigate how well these analytical copulas can approximate the overall dependence structure of empirical data. In addition, we compare our empirical copulas to the skewed Student's t-copula which is able to model asymmetries in the dependence structure. First, however, we shall have a closer look at the empirical asymmetry of the tail dependence itself.

\subsection{Asymmetry of the tail dependence}
\label{sec:asymmetry}
We are now going to elaborate on the features of the empirical copula densities. Their essential characteristic is the asymmetry between the \emph{lower-lower} and the \emph{upper-upper} corner: 
The dependence between large negative returns is always stronger than the one between large positive returns. Consequently, we observe a stronger dependence between coinciding downside movements than between coinciding upside movements. In other words, we distinguish an asymmetry between the bearish and the bullish market. This asymmetry is particularly evident for the original returns. \cite{Ang2002}, studying the dependence between US stocks and the US market, observed the same asymmetry on a correlation level.

\begin{figure}[htbp]
	\centering
		\includegraphics[width=0.49\textwidth]{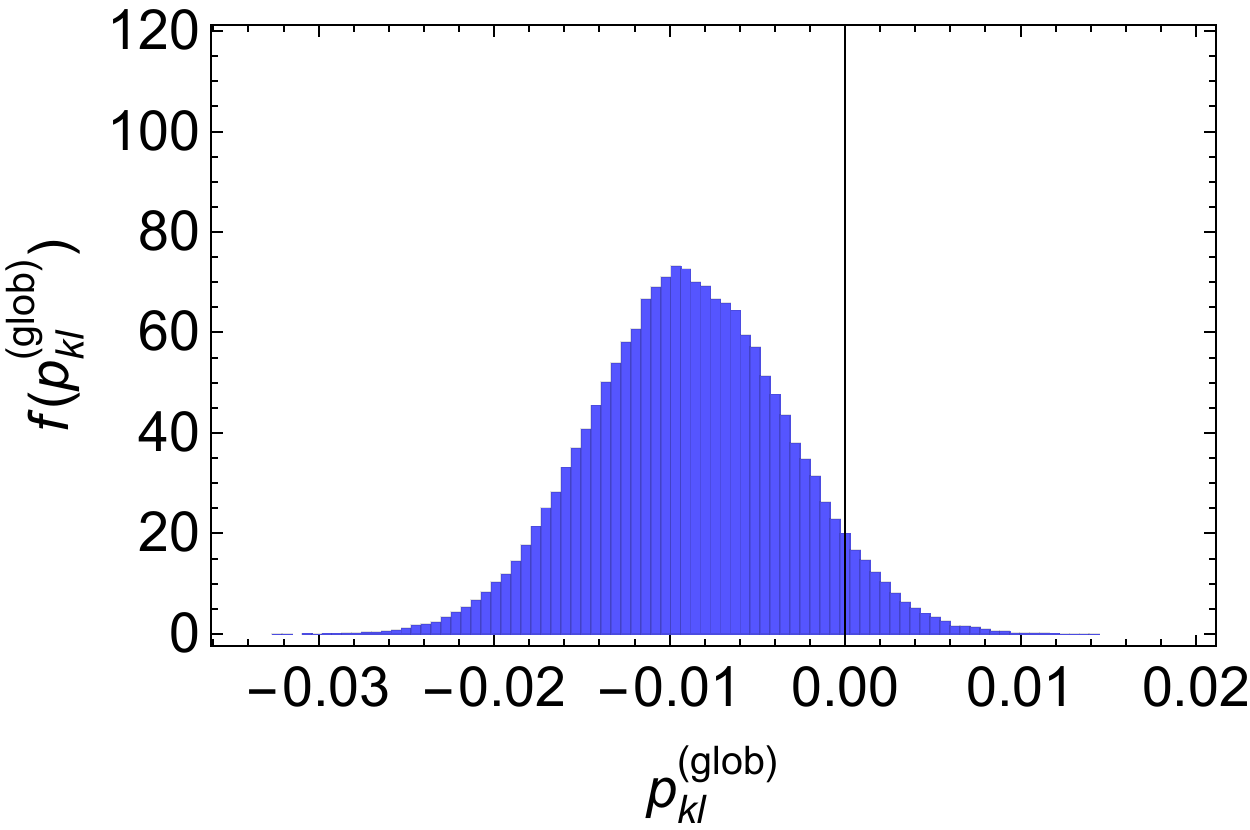}
		\includegraphics[width=0.49\textwidth]{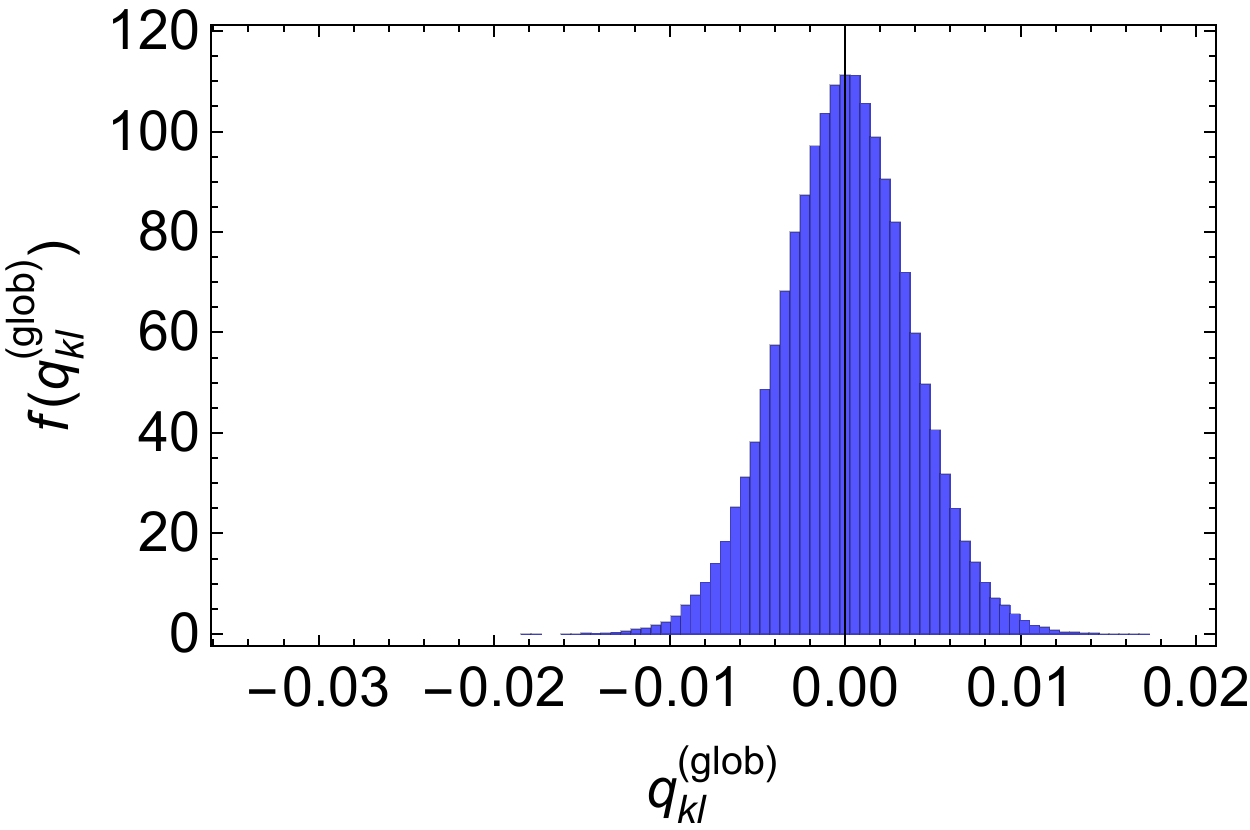}
		\includegraphics[width=0.49\textwidth]{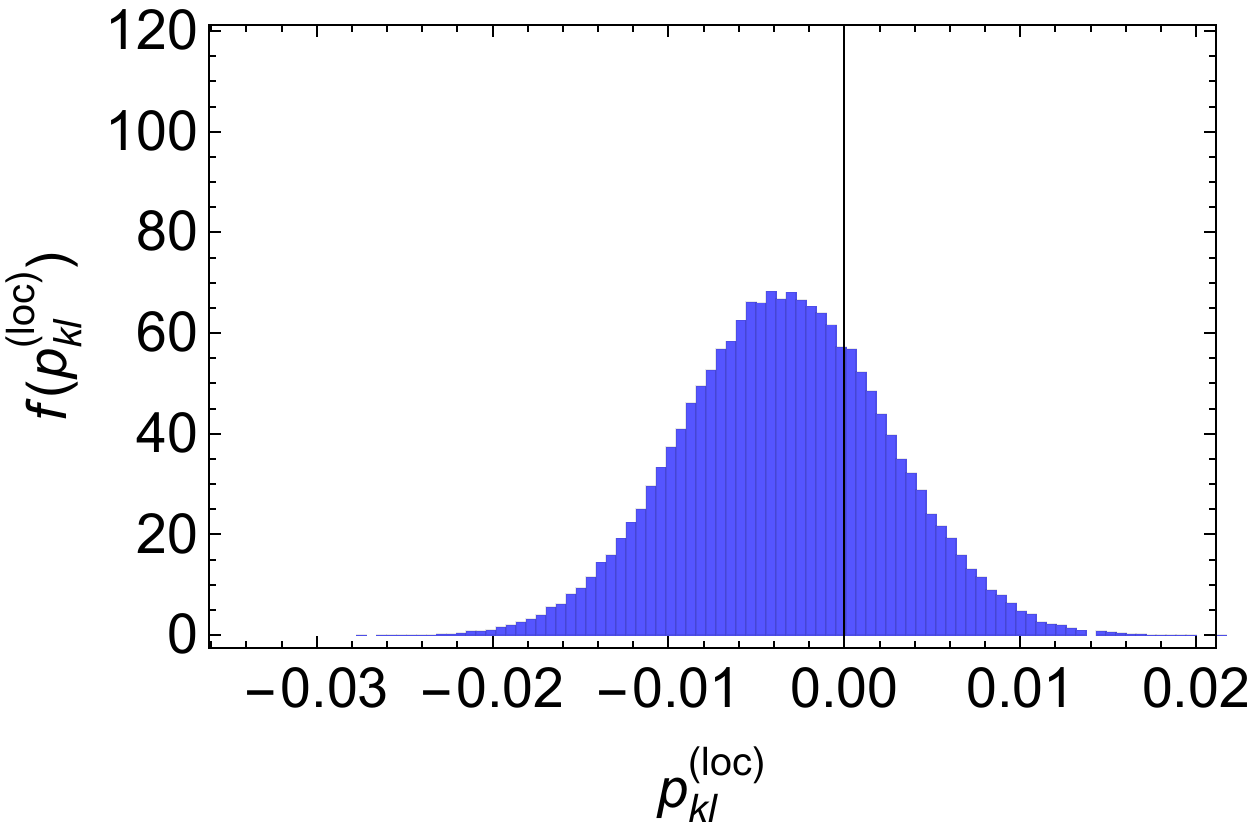}
		\includegraphics[width=0.49\textwidth]{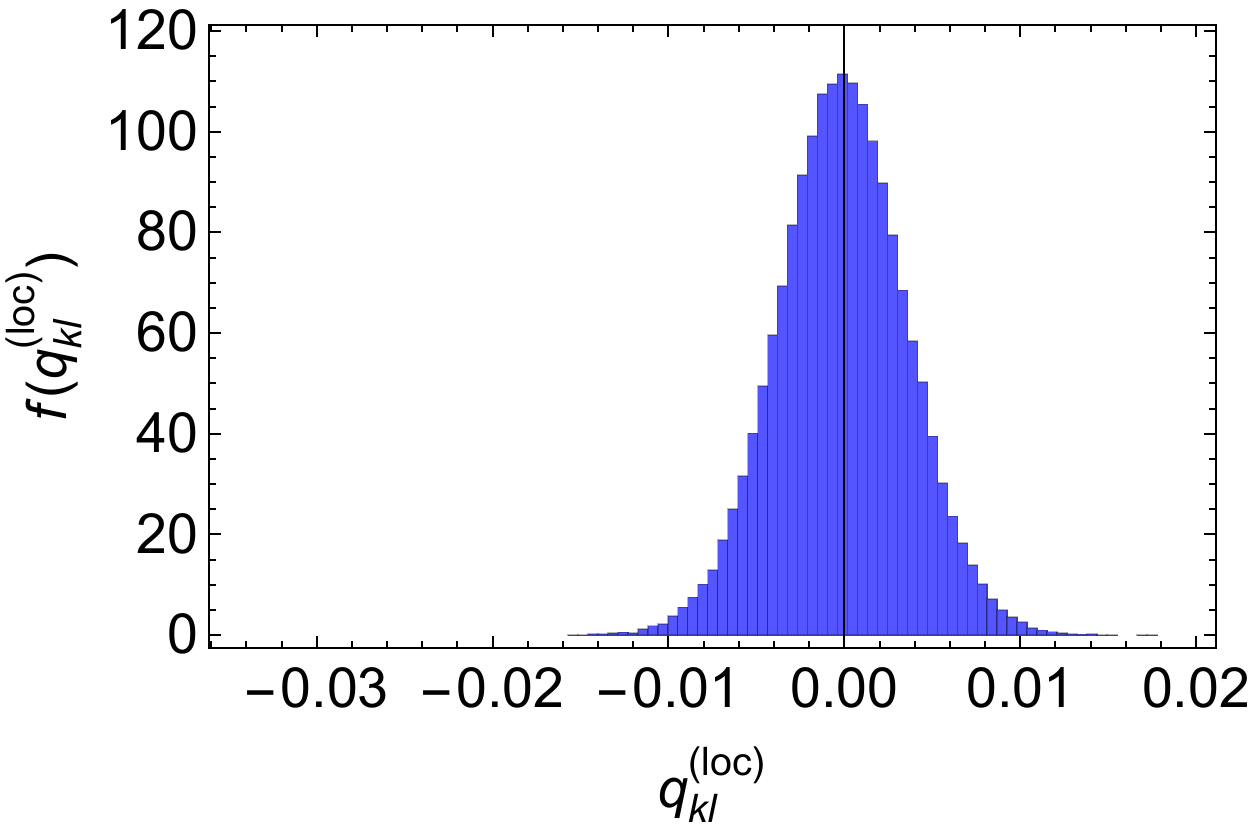}
	\caption{Histograms of the tail dependence asymmetries. Left: asymmetry for the positive tail dependence, $p_{k,l}$, right: asymmetry for the negative tail dependence, $q_{k,l}$. Top: for original returns, bottom: for locally normalized returns.}
	\label{fig:histdiff}
\end{figure}

To quantify this empirical asymmetry, we estimate the tail dependence via integrating over the 0.2$\times$0.2-area in all four corners for each of the $K(K-1)/2$ empirical pairwise copula densities. We receive the tail dependence asymmetries by subtracting the integrated areas in the opposite corners of the empirical copula densities:
\begin{align}
p_{k,l} &= \int\limits_{0.8}^{1} \mathrm du\int\limits_{0.8}^{1} \mathrm dv\,\cop_{k,l}(u,v) - \int\limits_{0}^{0.2} \mathrm du\int\limits_{0}^{0.2} \mathrm dv\,\cop_{k,l}(u,v)
\label{eq:postail} \\ 
q_{k,l} &= \int\limits_{0}^{0.2} \mathrm du\int\limits_{0.8}^{1} \mathrm dv\,\cop_{k,l}(u,v) - \int\limits_{0.8}^{1} \mathrm du\int\limits_{0}^{0.2} \mathrm dv\,\cop_{k,l}(u,v)
\label{eq:negtail}
\end{align}
We call $p_{k,l}$ the \emph{positive} tail dependence asymmetry and $q_{k,l}$ the \emph{negative} tail dependence asymmetry.
The histograms of these tail dependence asymmetries, $f(p_{k,l})$ and $f(q_{k,l})$, respectively, are shown in figure \ref{fig:histdiff} for both data sets. 
While the negative tail dependence asymmetries $q_{k,l}$ are centered around zero for both original and locally normalized returns, we observe a distinct negative offset for the positive tail dependence asymmetries $p_{k,l}$. Hence, on \emph{average} there is no asymmetry in the negative tail dependence, ie between coinciding large positive and large negative movements, but there is an asymmetry in the positive tail dependence, ie between coinciding large negative and coinciding large positive movements. 
We notice additionally that the locally normalized returns show a much weaker asymmetry in the positive tail dependence than the original returns. 
This empirical finding can be interpreted as follows:
For the original return time series, events in the tails of the distribution reflect periods of high volatility. And since high volatility often occurs simultaneously in most stocks, the tail dependence also reflects these periods in particular. 
In contrast, the locally normalized returns are stationary with a constant volatilit of one. Therefore, the tail dependence in this case reflect the average behavior over the entire time period.
Thus, our empirical results imply that the asymmetry in the tail dependence is non-stationary, but it is particularly strong in periods with high volatility.
In fact, periods with high volatility coincide both with strong correlations and strong asymmetry in the dependence structure.

\subsection{Comparison to the Gaussian copula}
\label{sec:avggausscop}
\begin{figure}[htbp]
	\centering
		\includegraphics[width=0.49\textwidth]{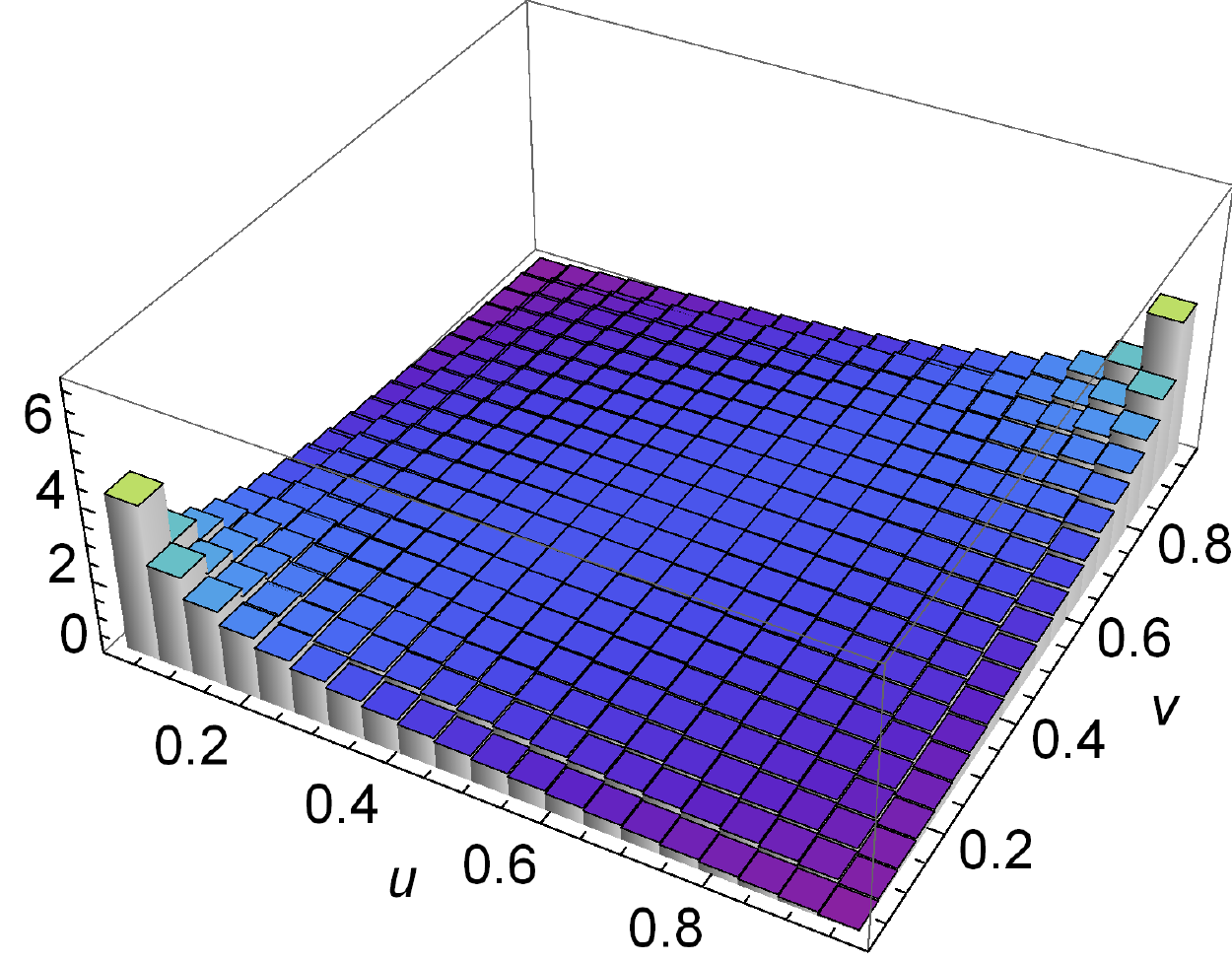}
		\includegraphics[width=0.49\textwidth]{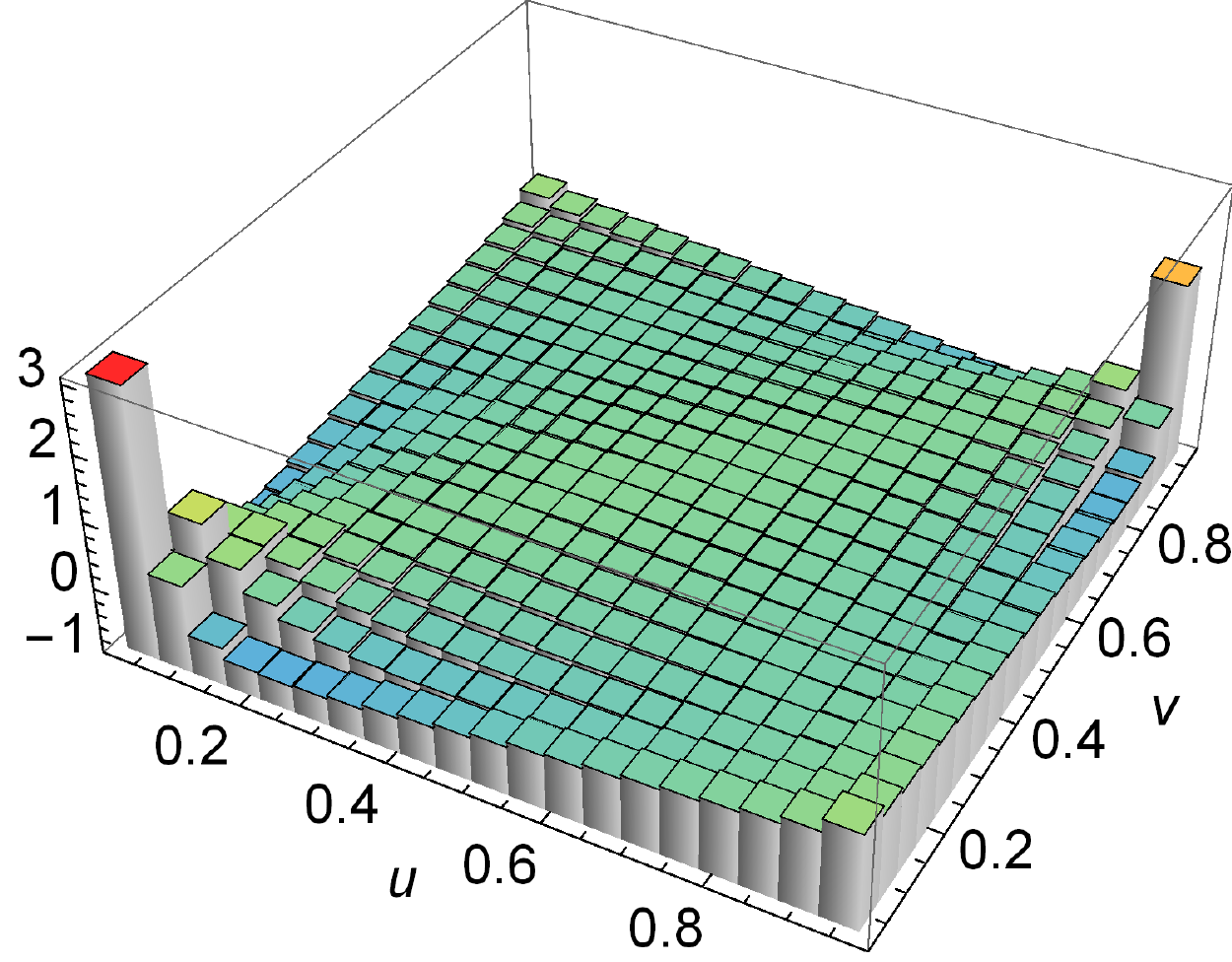}
		\includegraphics[width=0.49\textwidth]{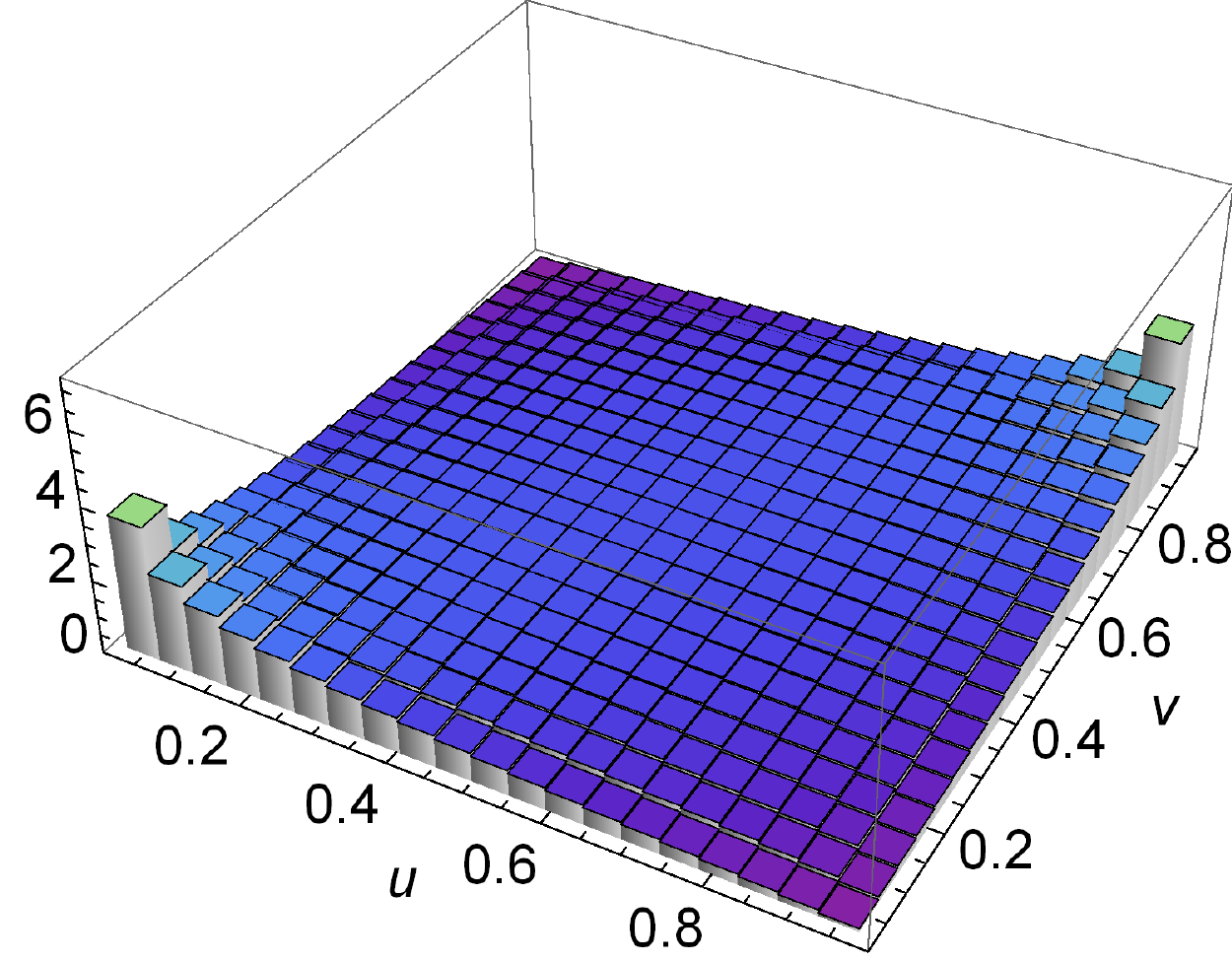}
		\includegraphics[width=0.49\textwidth]{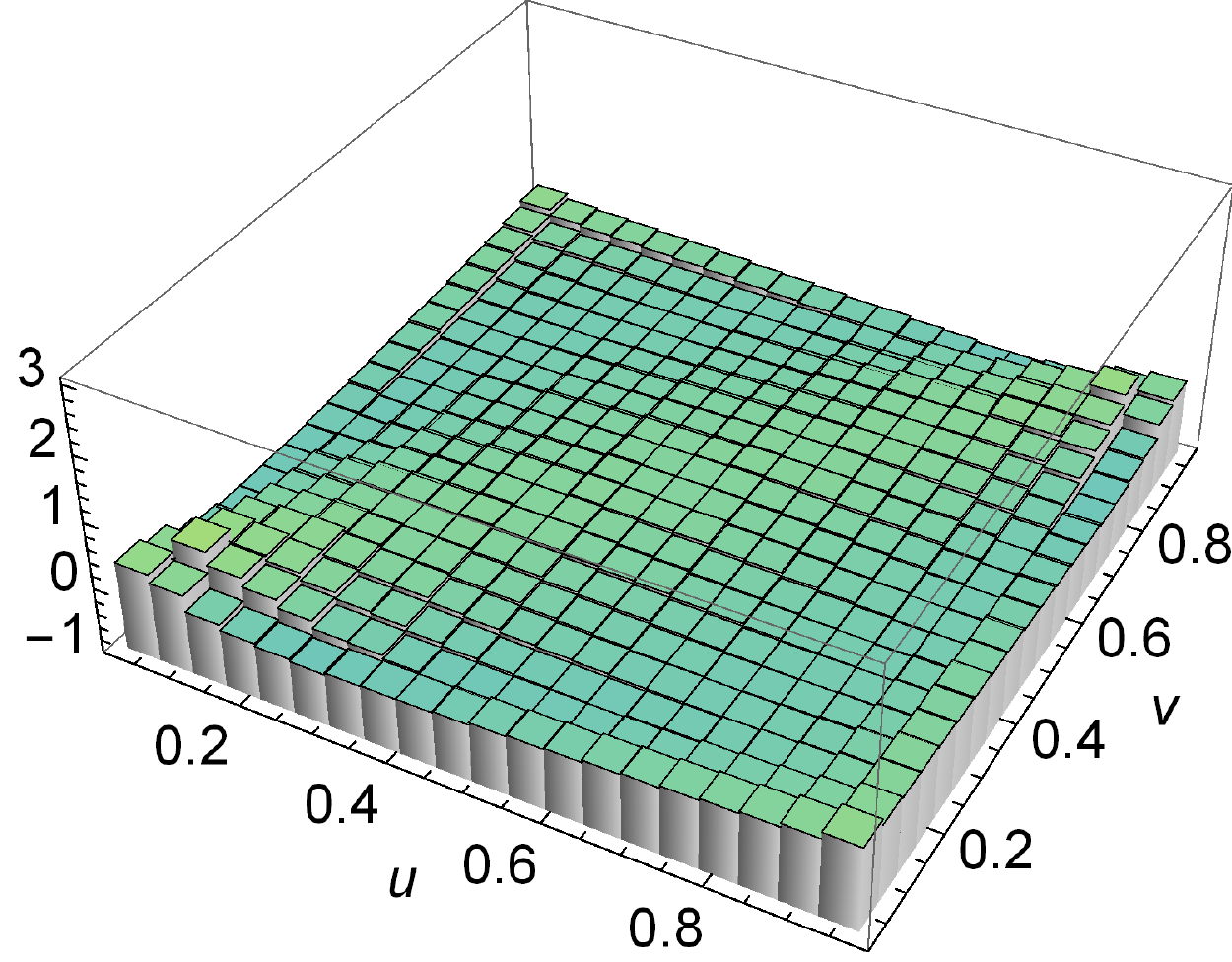}
	\caption{
Left: Gaussian copula densities $\cop_{c}^\text{G}(u,v)$, right: differences between the empirical copula density and the Gaussian copula density, $\cop^\text{(glob)}(u,v) - \cop_{c}^\text G(u,v)$ and $\cop^\text{(loc)}(u,v) - \cop_{c}^\text G(u,v)$, respectively. Top: for original returns, $\overline{c}^\text{(glob)}=0.44$, bottom: for locally normalized returns, $\overline{c}^\text{(loc)}=0.39$.
}
	\label{fig:gausscop}
\end{figure}

At first sight, our empirical copula densities seem to roughly resemble the Gaussian copula.
How good is this agreement quantitatively? 
In figure \ref{fig:gausscop} we compare the empirical copula densities with the respective analytical Gaussian copula density $\cop_{c}^\text{G}(u,v)$. Here we set the correlation coefficient in each Gaussian copula to the empirical average correlations: 
$\overline{c}^\text{(glob)}=0.44$ for original returns and $\overline{c}^\text{(loc)}=0.39$ for locally normalized returns.
In figure \ref{fig:gausscop} we observe clear deviations from the respective Gaussian copula densities. This is to be expected since the empirical copula densities are in fact an average over $K(K-1)/2$ pairwise copula densities with \emph{different} correlations. Hence, the comparison to a \emph{single} Gaussian copula with average correlation cannot yield suitable results. 
Indeed, for the original returns we find considerable deviations not only in the corners, but over the entire dependence structure. For the locally normalized returns, the corners are rather well described. Nonetheless, deviations over the whole dependence structure are clearly visible.
Table~\ref{tab:mse} summarizes the deviations between empirical and analytical copula densities in the form of least mean squares. For the original returns we find a least mean square of 27.52, while the locally normalized returns yield a smaller 5.26. 

\subsection{Comparison to a correlation-weighted Gaussian copula}
\label{sec:corrweigausscop}
\begin{figure}[htbp]
	\centering
		\includegraphics[width=0.49\textwidth]{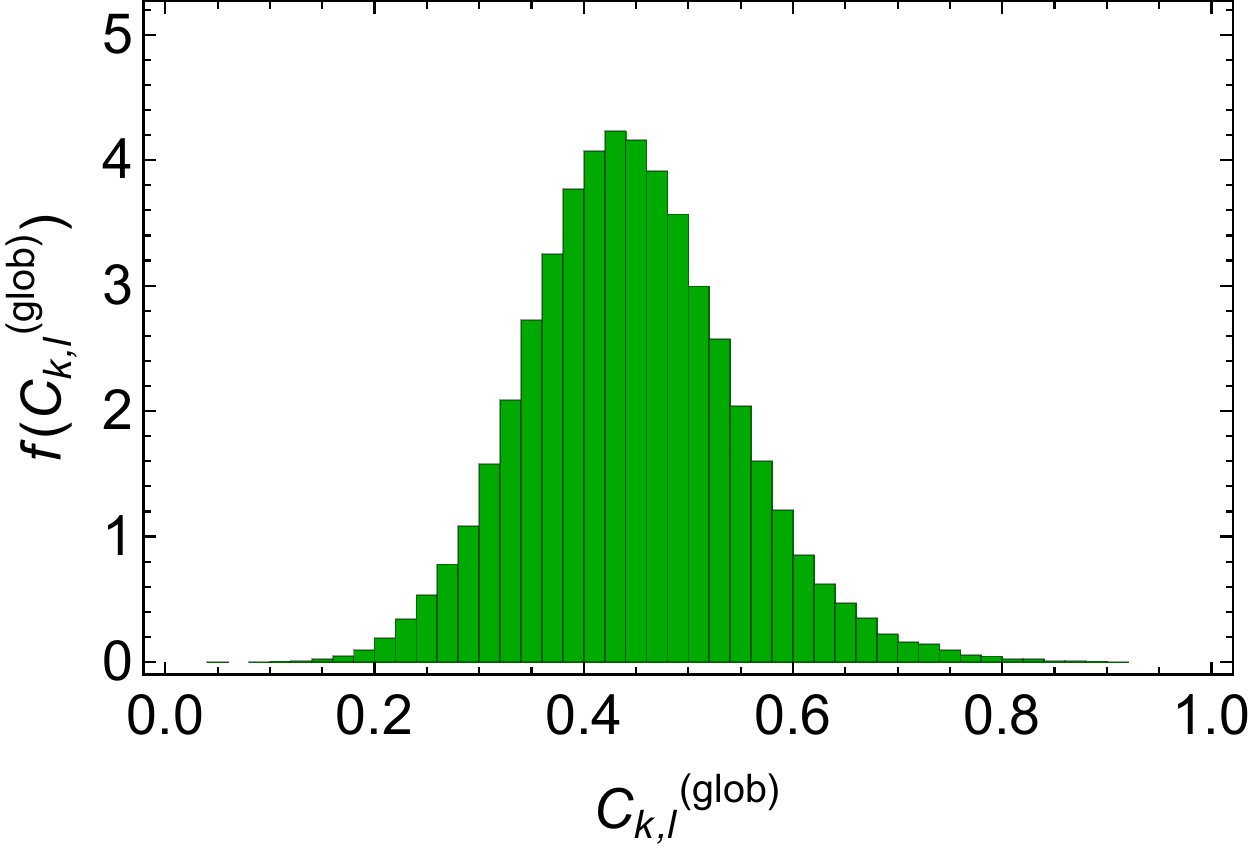}
		\includegraphics[width=0.49\textwidth]{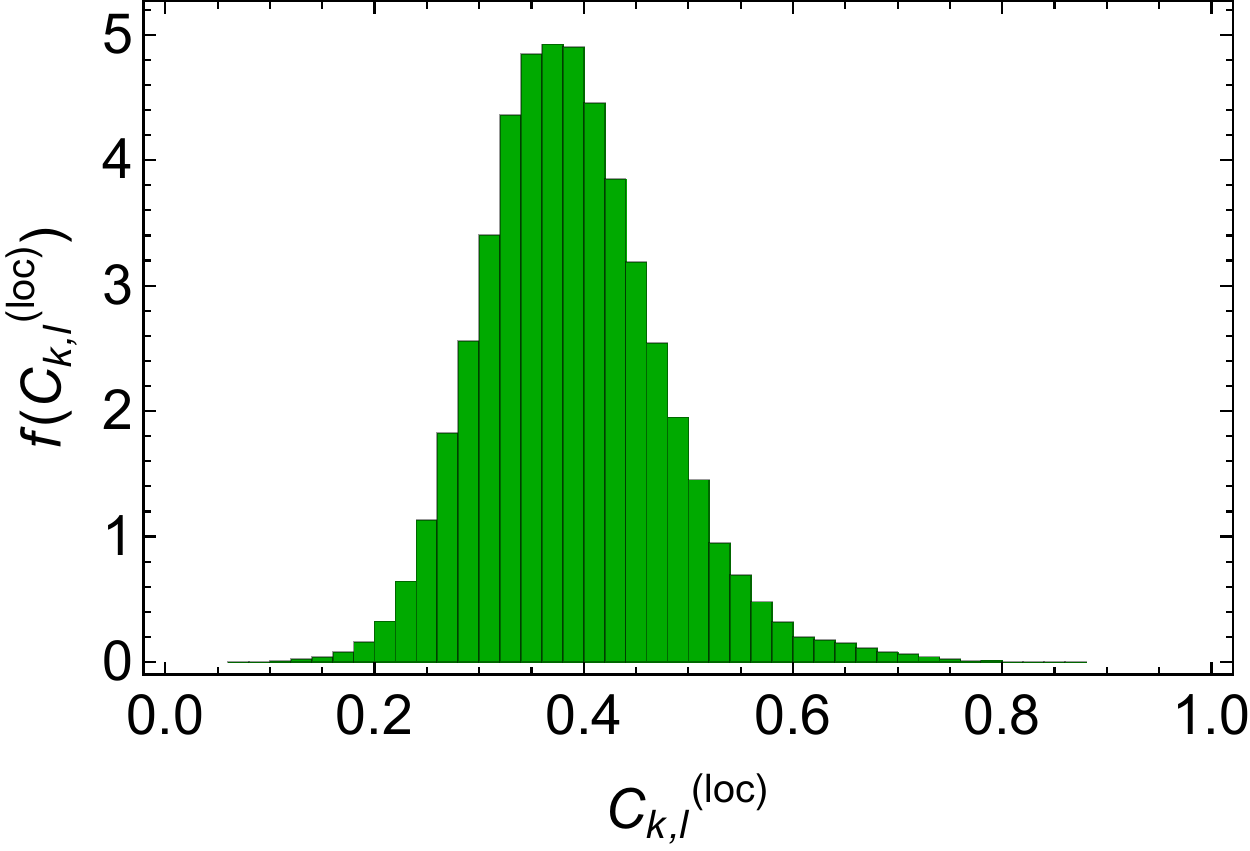}
	\caption{Histograms of correlations, left: for original returns, $f(C_{k,l}^\text{(glob)})$, right: for locally normalized returns, $f(C_{k,l}^\text{(loc)})$.}
	\label{fig:histcorr}
\end{figure}

\begin{figure}[htbp]
	\centering
		\includegraphics[width=0.49\textwidth]{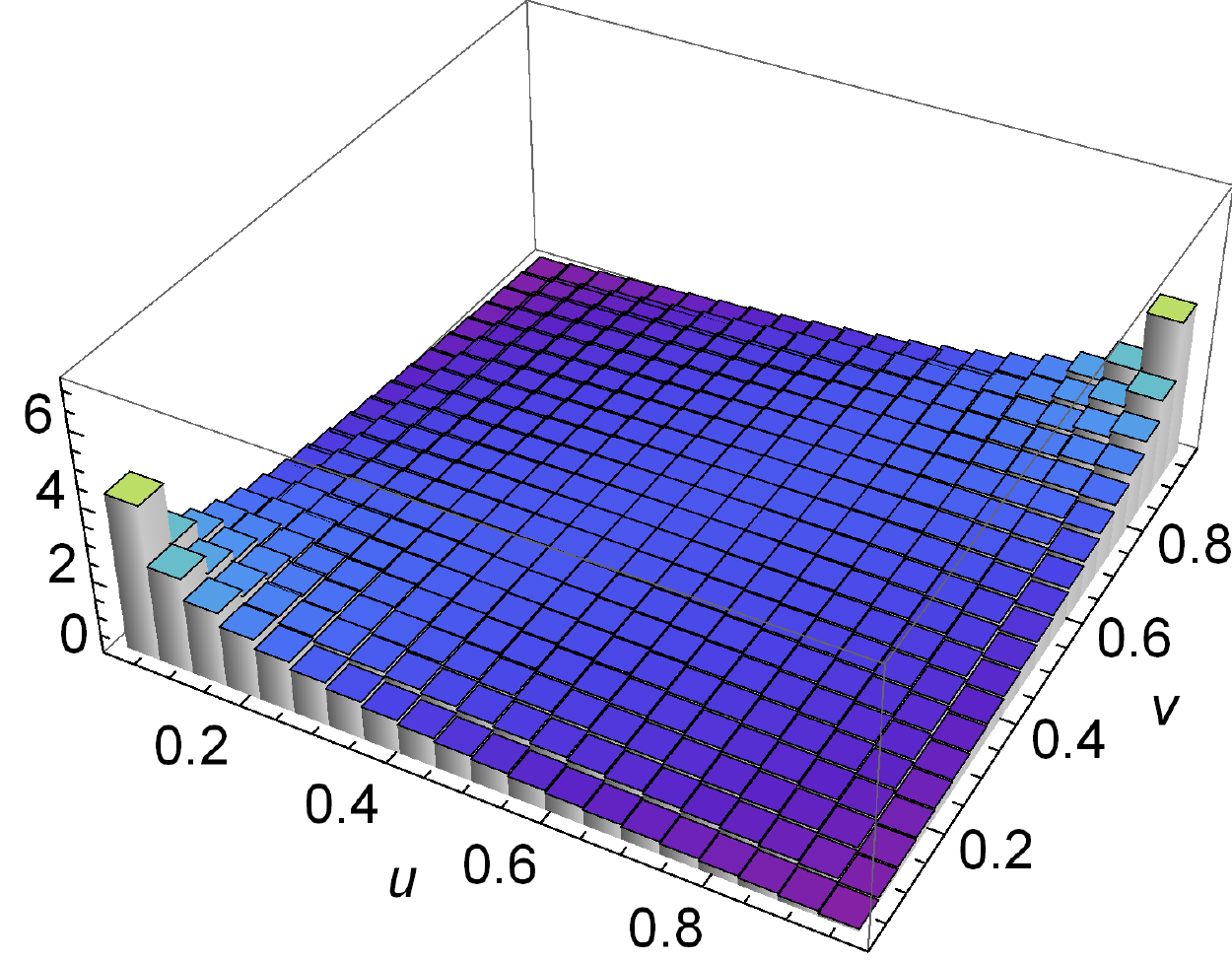}
		\includegraphics[width=0.49\textwidth]{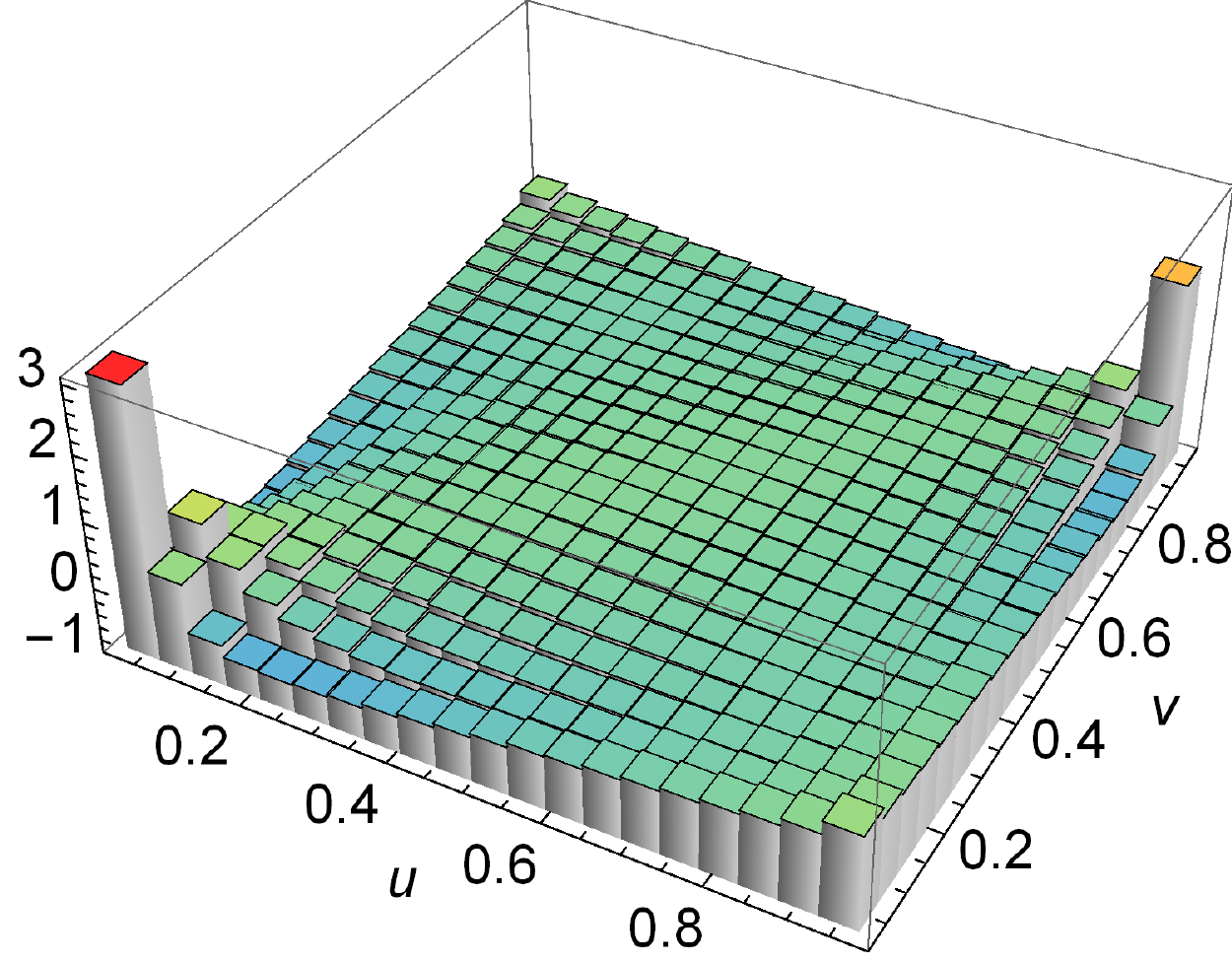}
		\includegraphics[width=0.49\textwidth]{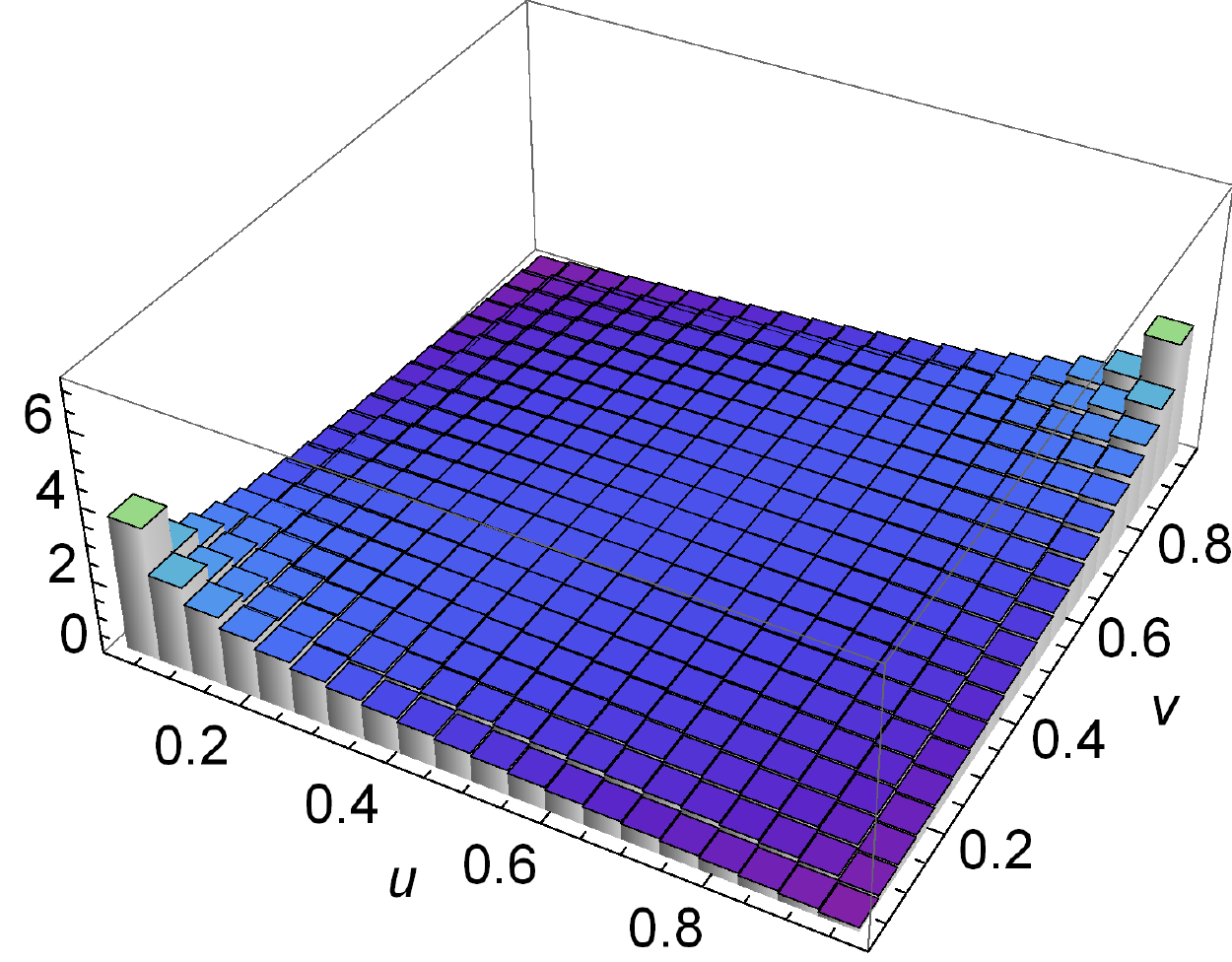}
		\includegraphics[width=0.49\textwidth]{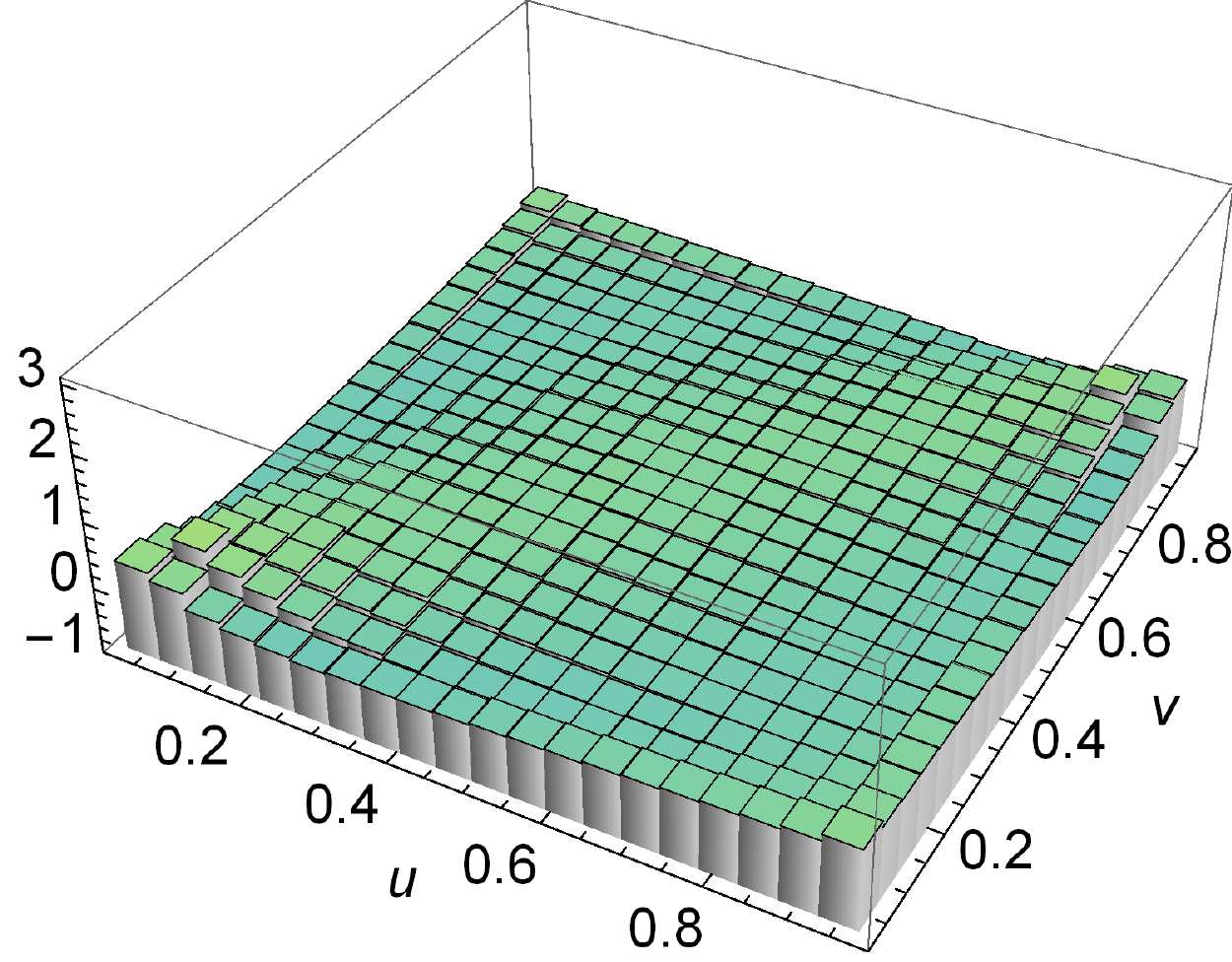}
	\caption{Left: correlation-weighted Gaussian copula densities $\cop(u,v)^\text{CWG}$, right: differences between the empirical copula density and the correlation-weighted Gaussian copula density, $\cop^\text{(glob)}(u,v) - \cop(u,v)^\text{CWG}$ and $\cop^\text{(loc)}(u,v) - \cop(u,v)^\text{CWG}$, respectively. Top: for original returns, bottom: for locally normalized returns.}
	\label{fig:weicorravggausscop}
\end{figure}

Aiming for a more suitable description of the empirical copulas, we are now considering a  weighted average of Gaussian copulas for different correlation coefficients. This takes into account the empirical average over different stock pairs, see equation \eqref{equ:avgempcop}. Figure \ref{fig:histcorr} shows the histograms of empirical correlation coefficients, $f(C_{k,l})$, for original and locally normalized returns estimated on the entire time series. The bin size is $\Delta c=0.02$. In our data sets we only observe positive correlations. 
We now obtain a Gaussian mixture by weighting each Gaussian copula density with the value of the probability density function of the respective correlation, $h(C_{k,l})=f(C_{k,l})\,\Delta c$. This yields the correlation-weighted Gaussian copula density
\begin{align}
\cop^\text{CWG}(u,v) = \sum_{C_{k,l}=-1}^{1} h(C_{k,l})\,\cop_{C_{k,l}}^\text G(u,v)
\end{align}
which is compared to the empirical copula densities in figure \ref{fig:weicorravggausscop}.
For both data sets we find only slight improvement over the single Gaussian copula, as depicted in figure \ref{fig:gausscop}. The least mean squares are only slightly smaller as well, see table~\ref{tab:mse}.
This can be attributed to the fact that correlations also vary in time, see eg, \cite{Schaefer2010}, \cite{Muennix2012}. Thus, it is not sufficient to take the average over different correlations into account. For a better approximation of the empirical copula densities, the correlations would have to been estimated on shorter time intervals. This, however, leads to increasing estimation errors for shorter estimation intervals. Consequently, a reliable attainment of time-dependent correlations is problematic. This is where the K-copula discussed in section \ref{sec:RMT} comes into play.

\subsection{Comparison to the K-copula}
\label{sec:kcop}
We now consider the K-copula density $\cop_{c,N}^\text K(u,v)$, which takes inhomogeneous and time-varying correlations into account. Here, the fluctuations of correlations around their mean value $c$ are characterized by the free parameter $N$.  We calculate the mean correlation coefficient for the original returns and the locally normalized returns. As noted above, we find 
$\overline{c}^\text{(glob)}=0.44$ and $\overline{c}^\text{(loc)}=0.39$, respectively.
The free parameter $N$ is  
fitted to the empirical copula densities by the method of least squares. We find $N^\text{(glob)}=3.2$ for original returns and $N^\text{(loc)}=7.8$ for locally normalized returns.
The parameter values for $\overline{c}$ and $N$ are also summarized in table~\ref{tab:par}.
The lower value for the original returns reflects the fact that the locally normalized returns have a constant variance of one. Hence, there are smaller fluctuations in the covariances, which are in this case simply the correlations. In the K-copula, smaller fluctuations of the covariances or correlations are described by larger values of $N$.

\begin{figure}[htbp]
	\centering
		\includegraphics[width=0.49\textwidth]{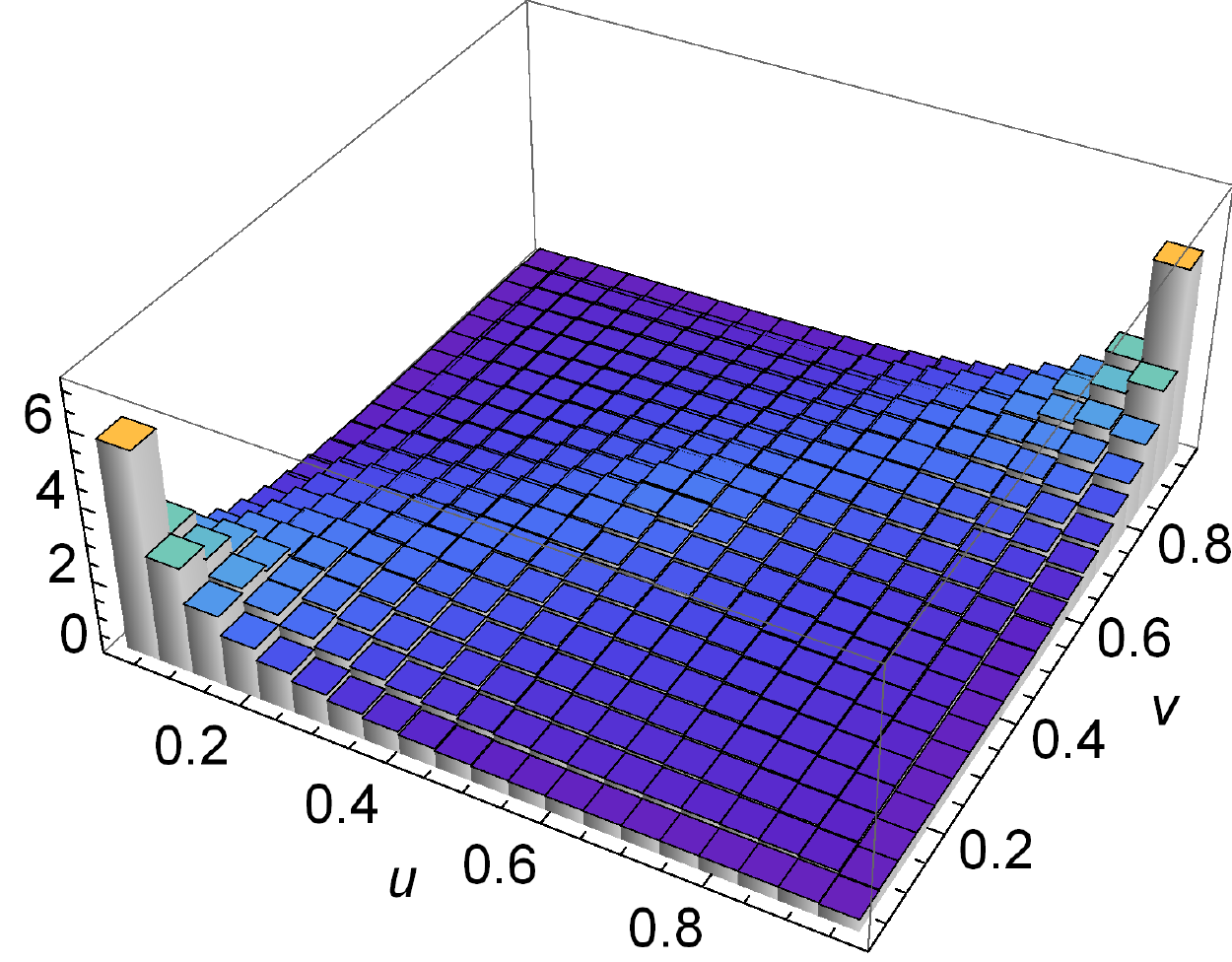}
		\includegraphics[width=0.49\textwidth]{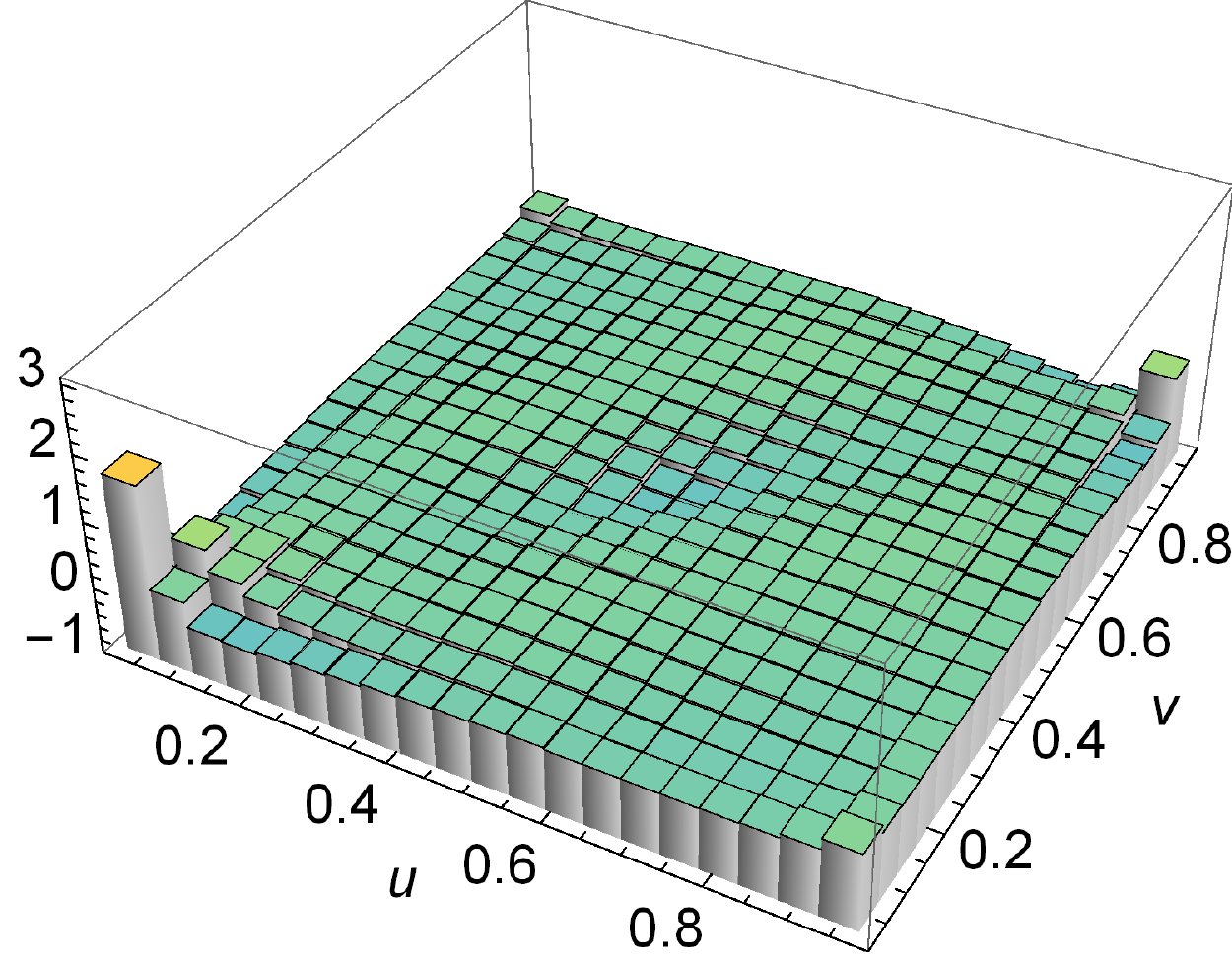}
		\includegraphics[width=0.49\textwidth]{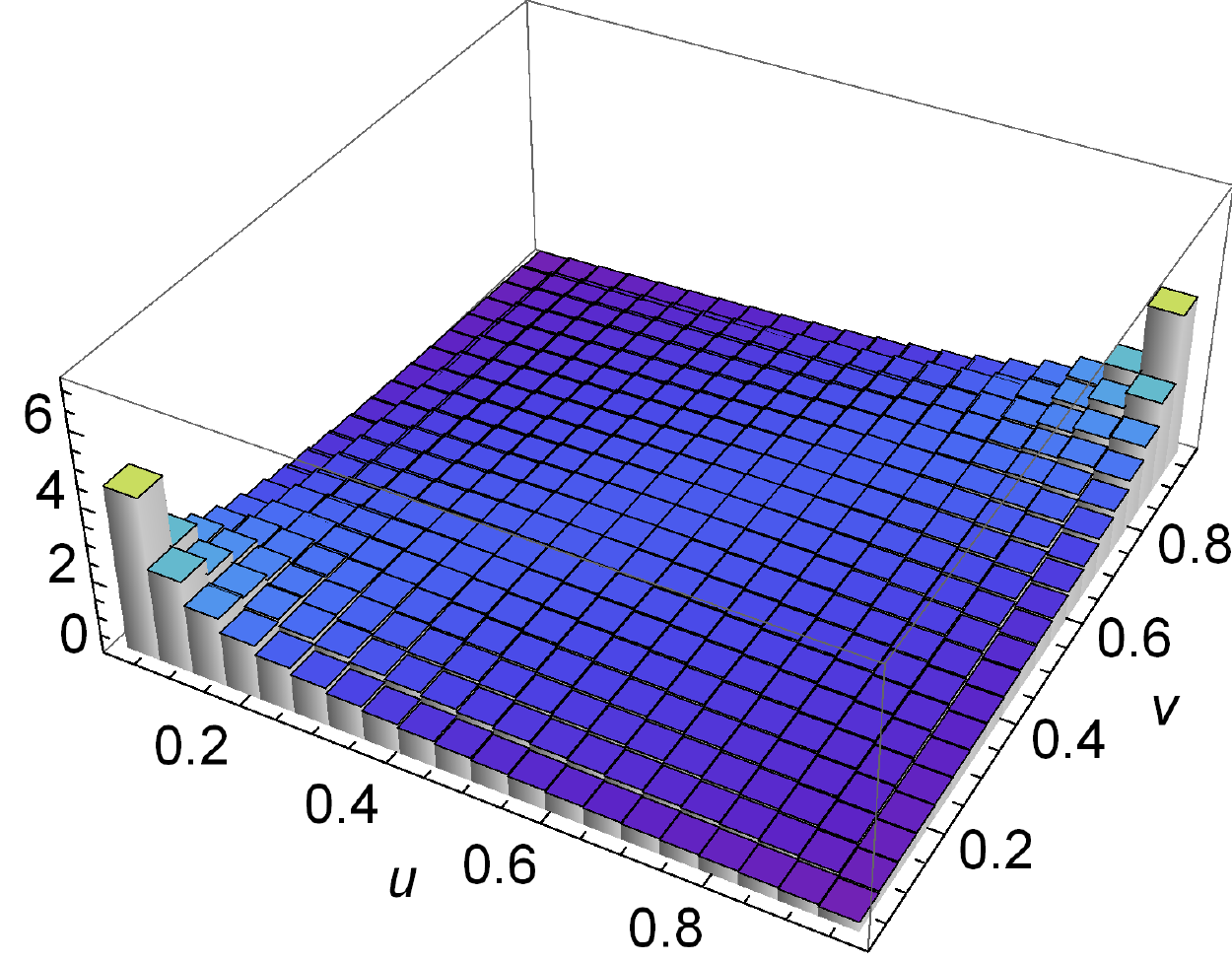}
		\includegraphics[width=0.49\textwidth]{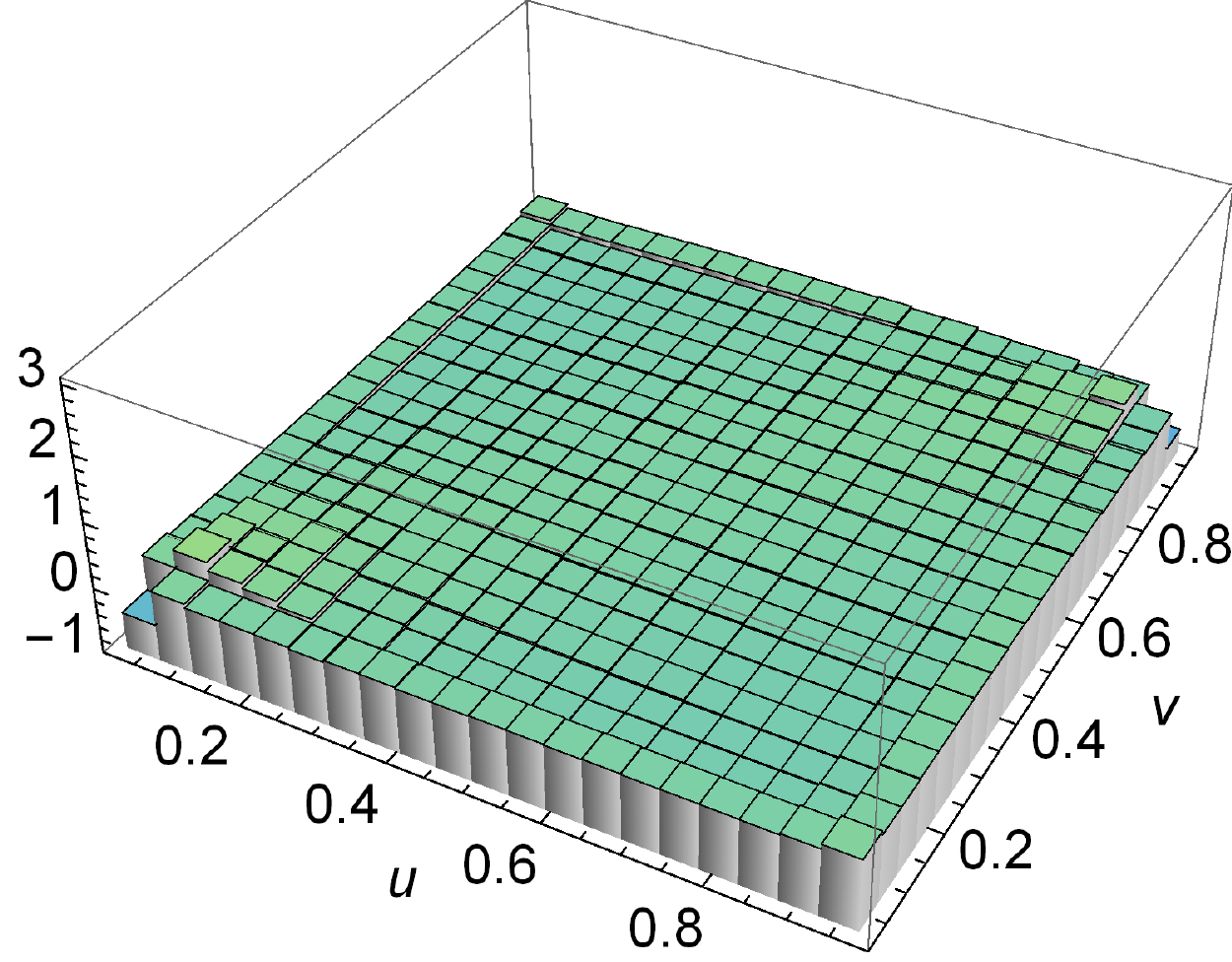}
	\caption{Left: K-copula densities $\cop_{c,N}^\text K(u,v)$, right: differences between the  empirical copula density and the K-copula density, $\cop^\text{(glob)}(u,v) - \cop_{c,N}^\text K(u,v)$ and $\cop^\text{(loc)}(u,v) - \cop_{c,N}^\text K(u,v)$, respectively. Top: for original returns, $\overline{c}^\text{(glob)}=0.44$, $N^\text{(glob)}=3.2$, bottom: for locally normalized returns $\overline{c}^\text{(loc)}=0.39$, $N^\text{(loc)}=7.8$.}
	\label{fig:copcorr}
\end{figure}

In figure \ref{fig:copcorr} the resulting K-copula densities are compared to the empirical copula densities. In both cases we find improved agreement with the empirical results.
This is also reflected in much lower values for the least mean squares, see table~\ref{tab:mse}.
However, for the original returns, a large deviation between empirical and analytical result remains. The overall structure of the empirical dependence is not captured very well by the K-copula.
For the locally normalized returns on the contrary, the K-copula yields a very good description of the empirical dependence. Only slight deviations persist in the lower-lower and upper-upper corner.
The asymmetry of the empirical copula densities cannot be captured by the K-copula density due to its symmetric character. This asymmetry is only weakly present in the case of locally normalized returns, which leads to a better agreement between K-copula and empirical copula.

\subsection{Comparison to the skewed Student's t-copula}
\label{sec:studcop}

The skewed Student's t-copula allows for an asymmetry in the dependence structure. 
Hence it is a natural candidate to compare our empirical findings with. Since we observe no asymmetry on average in the negative tail dependence of the empirical copula densities, see histograms in figure \ref{fig:histdiff} on the right, we choose the same value for both components of the skewness parameter vector $\boldsymbol\gamma$, $\gamma_1 = \gamma_2 = \gamma$. This leaves only an asymmetry in the positive tail dependence of the skewed Student's t-copula and reduces the number of parameters. The parameter values for $\nu$ and $\gamma$ are fitted to the empirical copula densities by the method of least squares, whereas the mean correlation $c$ is empirically determined by $\overline{c}^\text{(glob)}=0.44$ and $\overline{c}^\text{(loc)}=0.39$, respectively. We find $\nu^\text{(glob)}=3.3$, $\gamma^\text{(glob)}=0.06$ for original returns and $\nu^\text{(loc)}=8.0$, $\gamma^\text{(loc)}=0.04$ for locally normalized returns. The parameter values are summarized in table~\ref{tab:par2}.

\begin{figure}[htbp]
	\centering
		\includegraphics[width=0.49\textwidth]{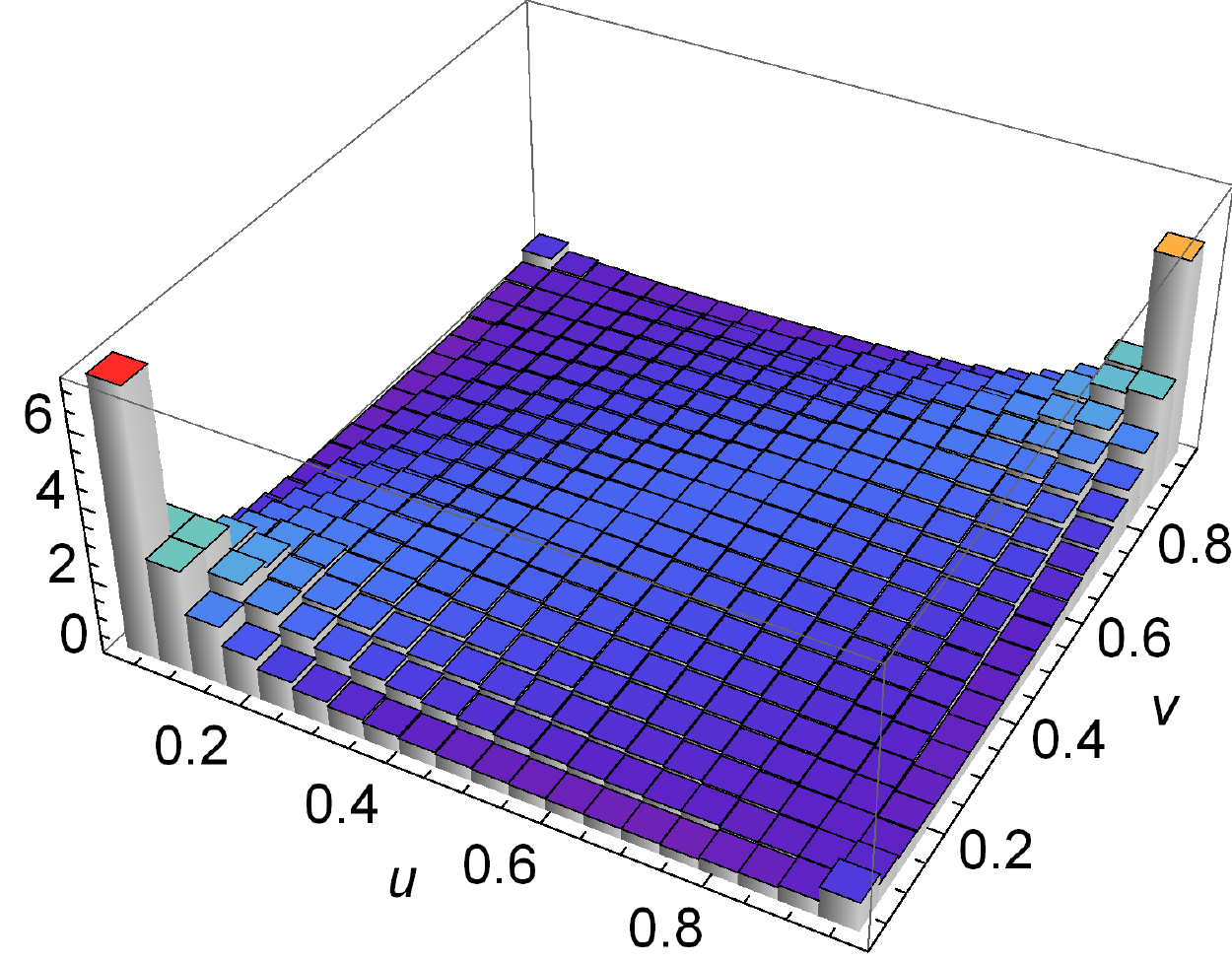}
		\includegraphics[width=0.49\textwidth]{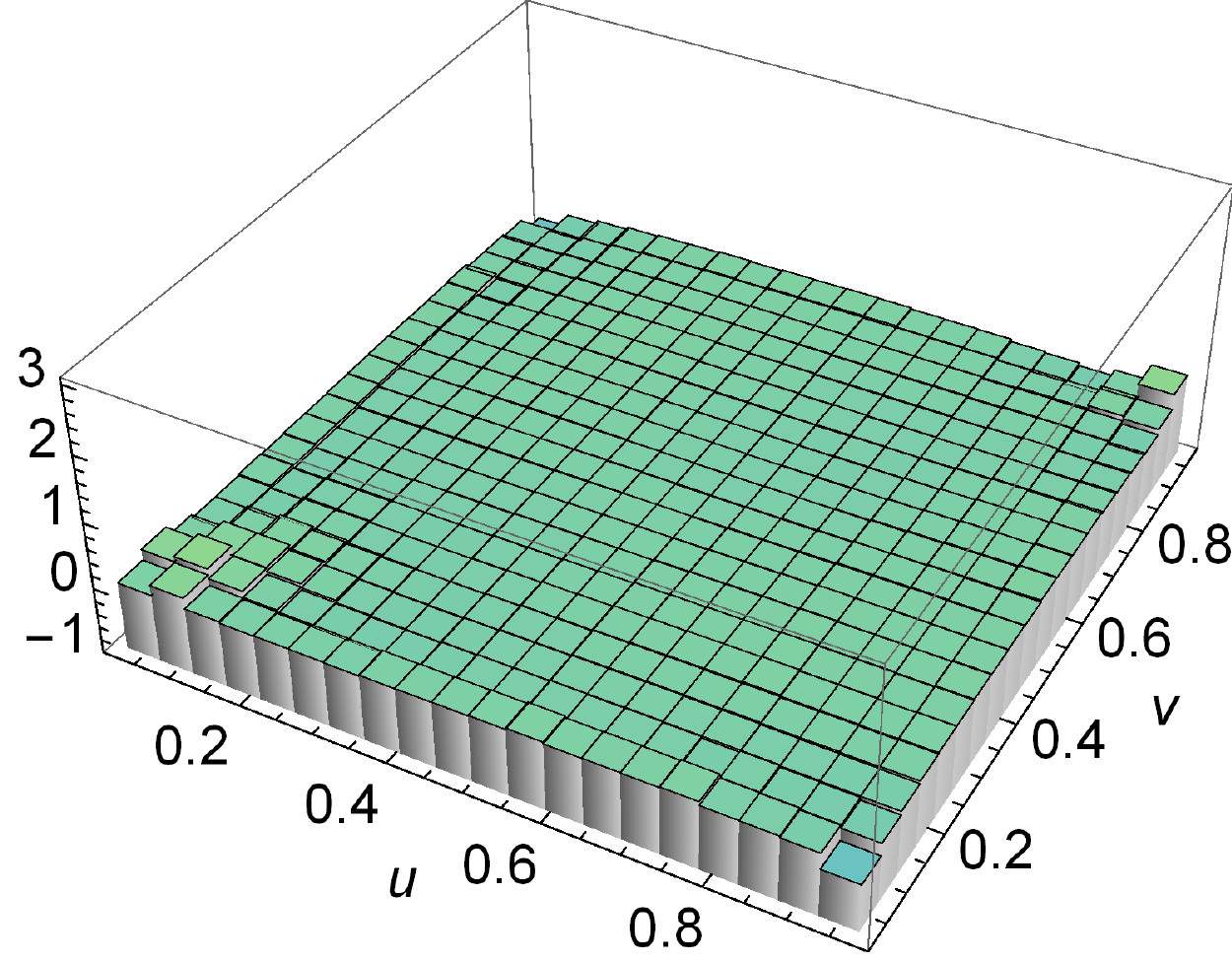}
		\includegraphics[width=0.49\textwidth]{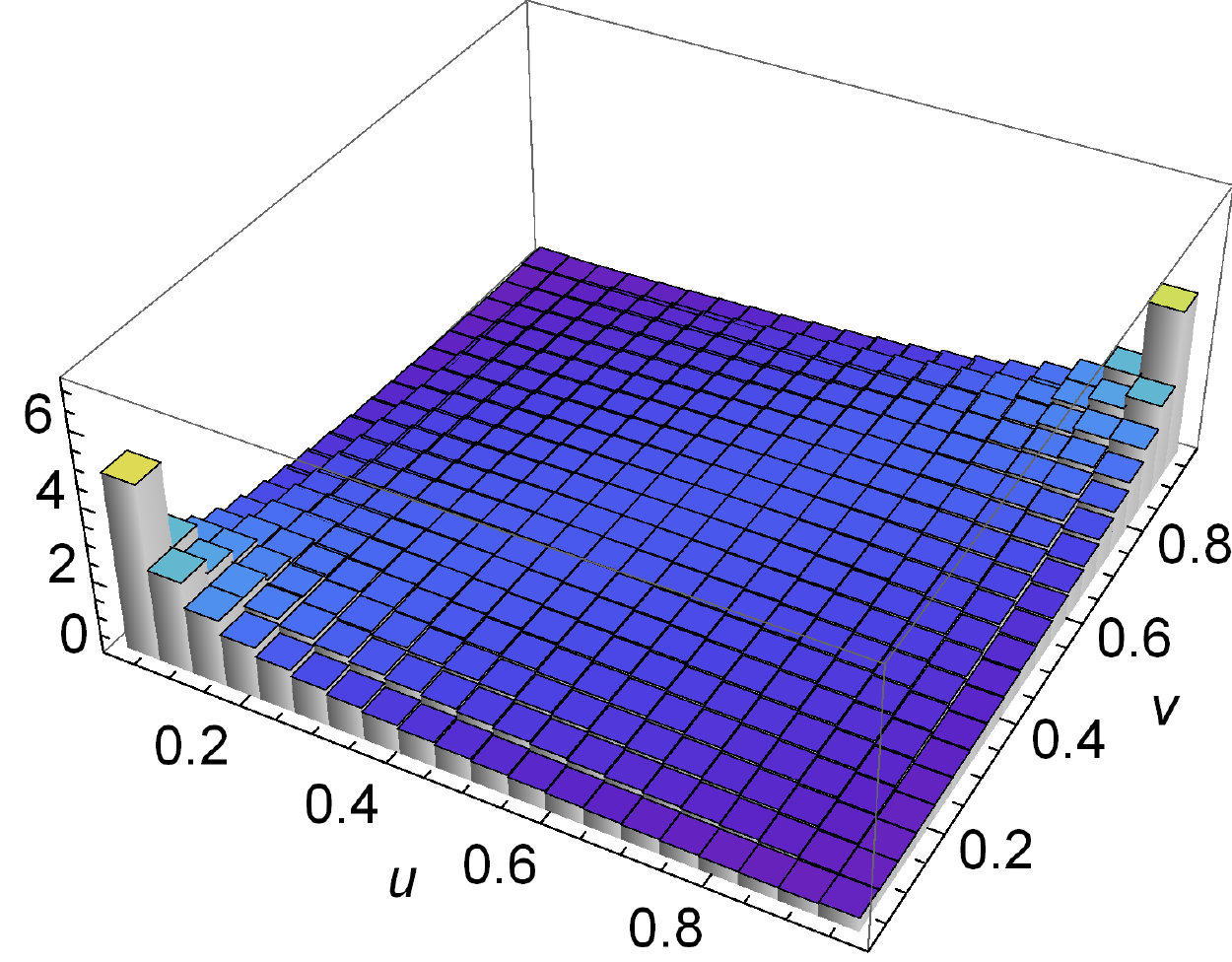}
		\includegraphics[width=0.49\textwidth]{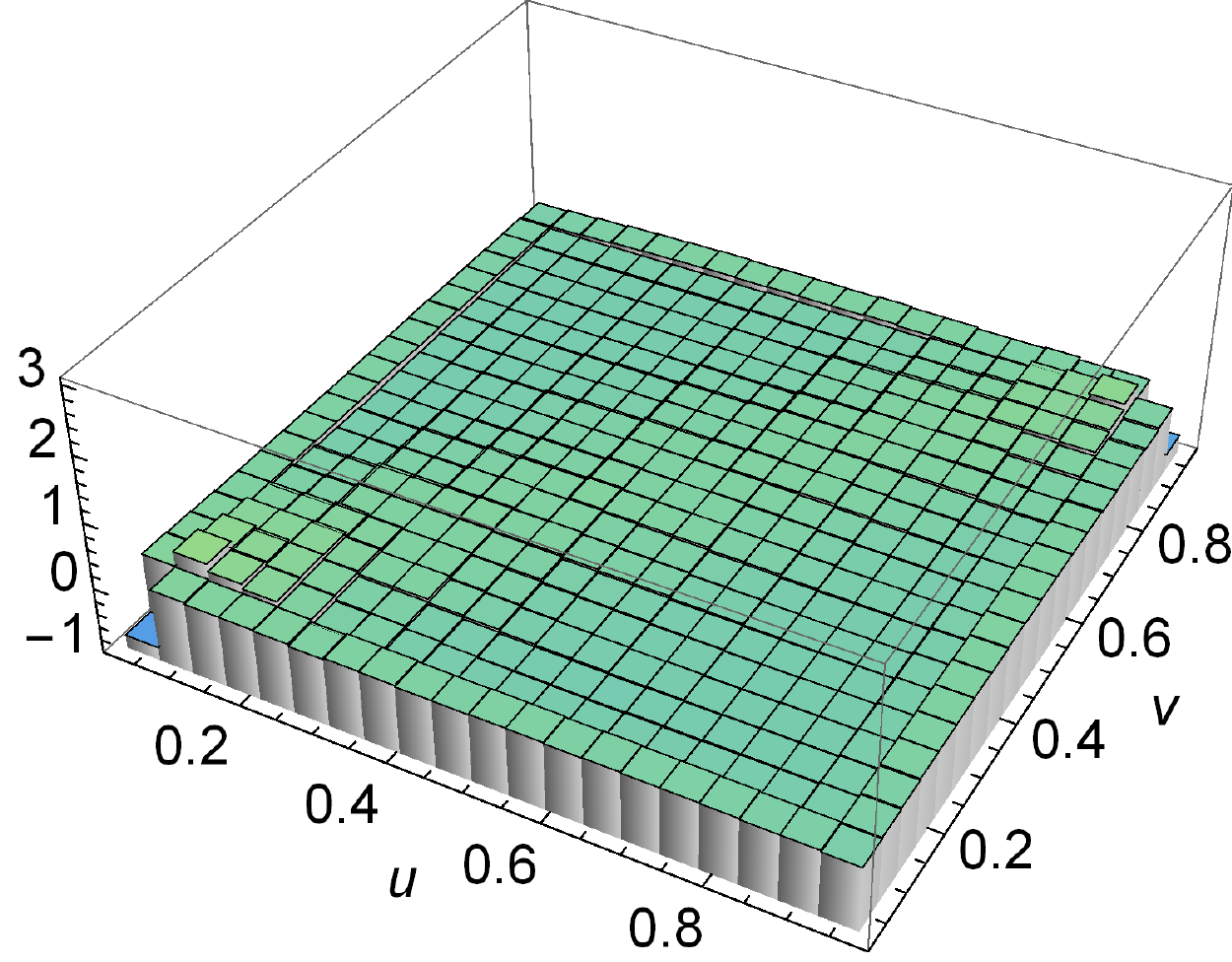}
	\caption{Left: Skewed Student's t-copula densities $\cop_{c,\nu,\gamma}^\text{t}(u,v)$, right: differences between the  empirical copula density and the skewed Student's t-copula density, $\cop^\text{(glob)}(u,v) - \cop_{c,\nu,\gamma}^\text{t}(u,v)$ and $\cop^\text{(loc)}(u,v) - \cop_{c,\nu,\gamma}^\text{t}(u,v)$, respectively. Top: for original returns, $\overline{c}^\text{(glob)}=0.44$, $\nu^\text{(glob)}=3.3$, $\gamma^\text{(glob)}=0.06$, bottom: for locally normalized returns, $\overline{c}^\text{(loc)}=0.39$, $\nu^\text{(loc)}=8.0$, $\gamma^\text{(loc)}=0.04$.}
	\label{fig:copstudt}
\end{figure}

Figure \ref{fig:copstudt} illustrates the resulting skewed Student's t-copula densities and their difference to the empirical copula densities. For original returns, ie the global scale, we find a remarkable agreement. The skewed Student's t-copula is able to capture the overall dependence structure very well, including the asymmetry in the positive tail dependence. There are only small deviations in the corners, in particular a mild overestimation for the negative tail dependence of extreme events.
For locally normalized returns, ie the local scale, the skewed Student's t-copula provides a slightly worse fit of the empirical data than the K-copula does. There is not much asymmetry to capture in the empirical copula density. Hence, the skewness parameter plays only a minor role in this case. Furthermore, the positive tail dependence is more pronounced for the skewed Student's t-copula than for the K-copula. Here, however, this is not reflected in the empirical dependence structure.

\begin{table*}[htbp]
\centering
\caption{Least mean squares between analytical and empirical copula densities.}
\label{tab:mse}
\begin{tabular}{rrrrrrrr}
\hline
Analytical density  & original returns  & loc.~normalized\\[0.5ex] 
\hline
Gaussian copula & 27.52 & 5.26 \\
corr.-weighted Gaussian copula & 26.42 & 5.11 \\
K-copula & 6.26 & 2.27 \\ 
skewed Student's t-copula & 0.65 & 2.87 \\ 
\hline
\end{tabular}
\end{table*}

\begin{table*}[htbp]
\centering
\caption{Parameter values for the K-copula densities.}
\label{tab:par}
\begin{tabular}{rrrrrrrr}
\hline
Parameter  & original returns & loc.~normalized\\[0.5ex] 
\hline
$\overline{c}$ & 0.44 & 0.39 \\
$N$ & $3.2$ & $7.8$ \\ 
\hline
\end{tabular}
\end{table*}

\begin{table*}[htbp]
\centering
\caption{Parameter values for the skewed Student's t-copula densities.}
\label{tab:par2}
\begin{tabular}{rrrrrrrr}
\hline
Parameter  & original returns & loc.~normalized\\[0.5ex] 
\hline
$\overline{c}$ & 0.44 & 0.39 \\
$\nu$ & $3.3$ & $8.0$ \\ 
$\gamma$ & $0.06$ & $0.04$ \\ 
\hline
\end{tabular}
\end{table*}

\section{Conclusion}
\label{sec:conclusion}

We presented an empirical study on the statistical dependencies between daily stock returns.
To this end we estimated empirical pairwise copulas for the original returns and for locally normalized returns. Considering the former allows us to study the dependence structure on a global scale. In the latter case, the non-stationary characteristics of return time series, ie time-varying trends and volatilities, were removed. This provides us with the dependence structure on a local scale.
How far does the concept of Gaussian dependence carry in the light of our empirical findings?
To answer this question we compared the empirical results not only with a single Gaussian copula, but also with a correlation-weighted average of Gaussian copulas and with the K-copula. The latter arises from a random matrix approach which models time-varying covariances by a Wishart distribution. This yields a K-distribution which adequately describes multivariate returns. We derived the resulting K-copula for the bivariate case and found very good agreement with empirical pairwise dependencies of the locally normalized returns.
Thus, we arrive at a consistent picture within the random matrix model: The K-distribution is able to describe the tail-behavior of the marginal return distributions and the K-copula captures the overall empirical dependence structure. 
This implies that Gaussian statistics, and thus also a Gaussian dependence structure, provides a good description on a \emph{local} scale.
However, on a \emph{global} scale, ie when the empirical distribution function, which is involved in the estimation of the copula, is applied to the original return time series instead, we find rather significant deviations from a K-copula. In particular, we observe a pronounced asymmetry in the positive tail dependence. 
Therefore, we also compare our empirical findings with a model that explicitly allows for such an asymmetry, the skewed Student's t-copula. And indeed, we find a rather compelling agreement with the empirical dependence structure of original returns.
For the local scale, however, the empirical copula exhibits only a mild asymmetry and is overall better described by the K-copula. 
How can we understand this?
For the original returns, the tail dependence reflects periods with high volatility, while for the locally normalized returns all periods contribute equally to the tail dependence. Thus, our results imply that the asymmetry in the tail dependence is non-stationary, and it is stronger in periods with high volatility.

\section*{Acknowledgments}
We thank Desislava Chetalova for helpful discussions.


\end{document}